\documentclass[letterpaper,12pt]{article}
\usepackage[letterpaper, margin=1in]{geometry}
\usepackage[utf8]{inputenc}
\usepackage{verbatim,array,multicol,palatino,amsfonts,amsmath, subfig, psfrag}
\usepackage{chicago}
\usepackage{color}
\usepackage[pdftex]{graphicx}
\usepackage{epstopdf}
\usepackage{hyperref}
\usepackage{latexsym,amssymb,epsfig,graphics, mathtools,amsthm, dcolumn,titlesec, authblk}
\newtheorem{proposition}{Proposition}

\usepackage{setspace}
\usepackage[normalem]{ulem}

\DeclareMathOperator*{\argmin}{arg\,min}
\titleformat*{\section}{\large\bfseries}

\title{ \bf {Pyramid Quantile Regression}}
\date{}
\author{T. Rodrigues\footnote{School of Mathematics and Statistics, University of New South Wales, Sydney 2052 Australia.}\:\,\footnote{CAPES Foundation, Ministry of Education of Brazil, Bras\'{i}lia - DF 70040-020, Brazil} ,
 J.-L. Dortet-Bernadet\footnote{Institut de Recherche Math\'ematique Avanc\'ee, UMR 7501 CNRS, Universit\'e de Strasbourg, Strasbourg, France.}\:\:  and Y. Fan$^{*}$\footnote{Communicating Author: {\tt Y.Fan@unsw.edu.au}}}

\begin{document}
\baselineskip=1.5\baselineskip

\maketitle

%\vspace{-0.5cm}

\begin{abstract}
Quantile regression models provide a wide picture of the conditional distributions of the response variable by capturing the effect of the covariates at different quantile levels. In most applications, the parametric form of those conditional distributions is unknown and varies across the covariate space, so fitting the given quantile levels simultaneously without relying on parametric assumptions is crucial. In this work we propose a  Bayesian model for simultaneous linear quantile regression.  More specifically, we propose to model the conditional distributions  by using  random probability measures known as quantile pyramids. Unlike many existing approaches, our framework allows us to specify meaningful priors on the conditional distributions, whilst retaining the flexibility afforded by the nonparametric error distribution formulation. Simulation studies demonstrate the flexibility of the proposed approach in estimating diverse scenarios, generally outperforming other competitive methods. We also provide conditions for posterior consistency. The method is particularly promising for modelling the extremal quantiles. Applications to  extreme value analysis and in higher dimensions are also explored through real data examples.
\\


\end{abstract}

\section{Introduction}

Since the seminal work by \citeN{KoenkerBasset1978} linear quantile regression has been recognized in recent years as a robust statistical procedure that offers a powerful and compelling alternative to ordinary linear mean regression. 
It has been successfully applied to a diverse range of fields whenever interest lies in the non-central parts of the response distribution, often found in the environmental sciences, medicine, engineering and economics.
Let $\tau$, $0<\tau<1$, be a probability value and let ${\cal X}$ be a bounded subspace of $\mathbb{R}^P$, for an integer $P \geq 1$.
The linear $\tau$-th quantile regression model  specifies the conditional distribution of a real response variable $Y$ given the value $X={\bf x}$ of a $P$ dimensional vector of covariates   
\begin{eqnarray}
Y|{\bf x} & \sim & \beta_{\tau}^0+ {\bf x}' \beta_{\tau}+\epsilon ,
\end{eqnarray}
for some unkown coefficients  $\beta_{\tau}^0 \in \mathbb{R}$ and $\beta_{\tau} \in \mathbb{R}^{P}$, and for a noise variable $\epsilon$ whose $\tau$-th conditional quantile is $0$, {\it i.e.}  $Q_{\epsilon} (\tau|{\bf x}) \equiv \inf\{a: P(\epsilon \leq a |X={\bf x}) \geq \tau\} = 0$ or  $\mathbb{P}(\epsilon\leq 0|X={\bf x})=\tau$.
Equivalently, we can write the $\tau$-th quantile of the conditional distribution of $Y$ given $X={\bf x}$ as $Q_{Y}(\tau|{\bf x}) =  \beta_{\tau}^0+{\bf x}'\beta_{\tau}.$

Let $(y_i,{\bf x}_i)_{i=1,...,N}$ be $N$ observed values of $(Y,X)$. 
Frequentist \textcolor{black}{inference on the linear quantile regression model} typically leaves the noise distribution unspecified and  
the estimation of \textcolor{black}{$(\beta_{\tau}^0,\beta_{\tau})$} is carried out by solving the minimization problem, 
\begin{eqnarray*}
(\hat{\beta}_{\tau}^0,\hat{\beta}_{\tau})=\argmin_{ (\beta^0, \beta) } \sum_{i=1}^N \rho_{\tau}(y_i-\beta^0-{\bf x}_i'\beta) \, ,
\end{eqnarray*}
where the so-called ``check function'' $\rho_{\tau}(.)$ is given by $\rho_{\tau}(\epsilon)=\tau \epsilon $ if $\epsilon\geq 0$ and $\rho_{\tau}(\epsilon)=(\tau-1) \epsilon $ otherwise (see \citeNP{KoenkerBasset1978}).  \textcolor{black}{Inference is usually based on asymptotic arguments}, see \citeN{koenker2005} for more details and properties of this approach. Bayesian treatment of quantile regression is more challenging, since a specification of a likelihood can be problematic. In recent years, the asymmetric Laplace error model has emerged as a popular tool for Bayesian inference (\shortciteNP{YuMoyeed2001}), largely due to its flexibility and simplicity, and the fact that the corresponding maximum likelihood estimate is the solution of the minimization problem above. 
It was shown in  \shortciteN{sriramrg13} that, under mild conditions, the asymmetric Laplace can produce a Bayesian consistent posterior inference for the case of linear quantiles.   
However, in applications to real data,  we do not really expect the distribution of the underlying data to follow an asymmetric Laplace distribution.  
Empirically, several authors have demonstrated that the asymmetric Laplace model does not have good coverage probabilities, see for example \shortciteN{reichbondellw08}. Other authors have tried to model the error distribution flexibly with nonparametric distributions, constraining the $\tau$-th quantile of the error distribution to be zero. See {\it e.g.} \shortciteN{kottasg01}, \shortciteN{hanson2002}, \shortciteN{kottask09} or \shortciteN{reichbondellw08} who propose the use of various nonparametric distributions including infinite mixture of Gaussians, Dirichlet process mixtures and mixture of P\'{o}lya trees.

In many applications, quantile estimates at several different quantile levels are needed to provide a precise description of the conditional distribution.  A well known problem with separately fitted quantile regression planes is that they can cross, violating the definition of quantiles. A possible solution is to use a second stage adjustment to the initial fits, see for example \shortciteN{hall99},  \shortciteN{dettev08} and \shortciteN{chernozhukovfg09} in the frequentist setting, or more recently \shortciteN{rodrigues2015} in a Bayesian setting. 
Another possible solution is a   joint estimation of multiple quantiles. This has been advocated by several authors, as it leads naturally to a greater borrowing of information across quantiles and a higher global efficiency for all quantiles of interest. Under this paradigm, \shortciteN{ReichFuentesDunson2011} proposed a model using Bernstein basis polynomials for spatial quantile regression. \shortciteN{tokdatk12} and \shortciteN{yang2015}  treat the regression coefficients as a function of $\tau$, using smooth monotone curves to model them under a Gaussian process prior. One of the common issues facing the more general modelling approach is that the likelihood is not available in analytic form, leading to the necessity to numerically approximate the likelihood values for each data observation. 
%This has the computational disadvantage of an added source of error from the numerical approximation, and sometimes prohibitive computational cost when the dataset is large, see for example \shortciteN{ReichFuentesDunson2011}.
A closed-form likelihood approach is proposed by \shortciteN{reichs13}, who extended the location scale model of \citeN{he97} to more flexibly model the quantile process. More recently, \shortciteN{fengch2015} proposed to use a linearly interpolated approximate likelihood derived from the quantiles, where the peudo-likelihood is available in analytical form, which approaches the true likelihood with increasing number of quantiles.

In this paper, we make the following contributions. First, we extend and modify the quantile pyramids described in \shortciteN{hjortw09} to the  regression setting, and we construct a flexible linear quantile model. Second, we show how meaningful priors can be placed directly on the quantiles, which can lead to better estimates. Third, we prove posterior consistency for the conditional quantiles. Finally, we provide an efficient method for parameter estimation via MCMC.

The article is organised as follows. In Section \ref{sec:pyramid} we recall the basic construction of the quantile pyramids studied in \shortciteN{hjortw09}. The proposed pyramid quantile regression (PQR) modelling is detailed in Section \ref{sec:regpyramid}, including its  theoretical properties and an estimation procedure. Extensive simulation studies are carried out in Section \ref{sec:sim}, where the proposed method is compared to the best alternative approaches. In Section \ref{sec:realex}, real examples illustrate PQR application to extreme quantile modelling and censored data analysis with a large number of covariates. The final section presents concluding discussions.

\section{Quantile pyramids for random distributions}\label{sec:pyramid}
Quantile pyramids was introduced by \shortciteN{hjortw09} as a method to define a random probability measure for nonparametric Bayesian inference. Contrary to the better known P\'{o}lya trees (\shortciteNP{ferguson74}, \shortciteNP{lavine1992}, \shortciteNP{lavine94}) that consider random probability masses and fixed partitions,  \shortciteN{hjortw09} propose the use of random quantiles with  fixed probabilities.

\begin{figure}[!ht]
    \centering
    \subfloat[\label{figa}]{\includegraphics[width=5.1cm, height=7cm, angle=0]{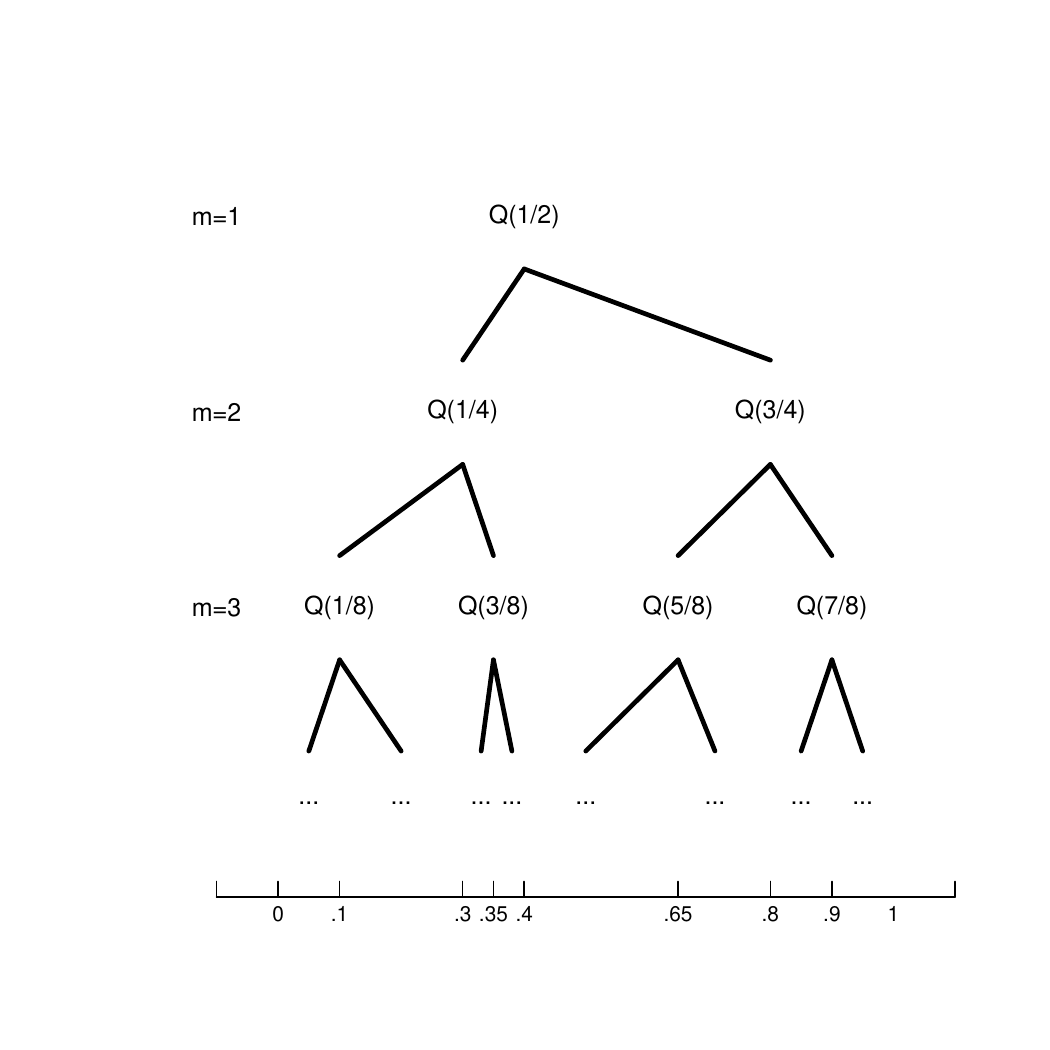}} \hspace{0.5cm}
    \subfloat[\label{figb}]{\includegraphics[width=5.1cm, height=7cm, angle=0]{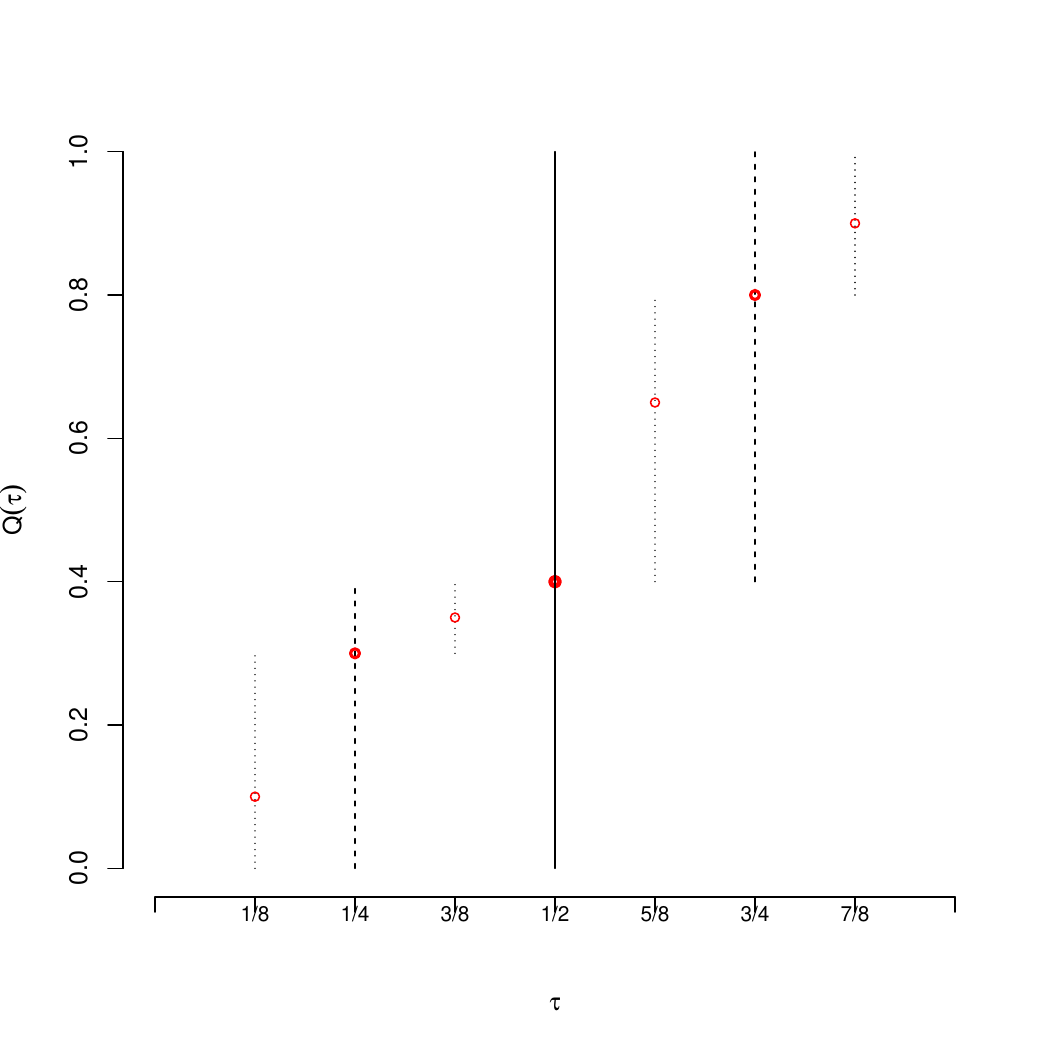}} \hspace{0.5cm}
    \subfloat[\label{figc}]{\includegraphics[width=5.1cm, height=7cm, angle=0]{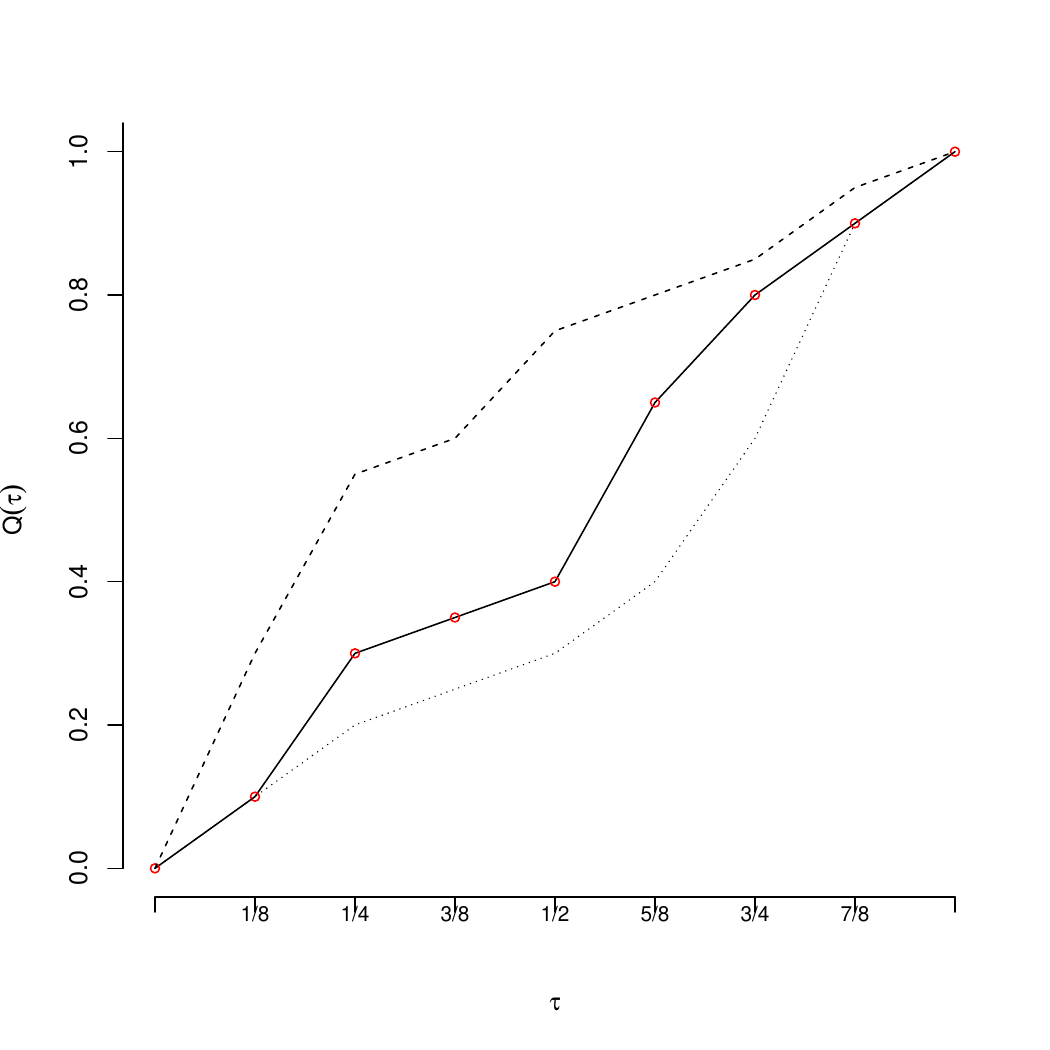}}
    \caption{The quantile pyramid generating process. Figure (a) shows the binary tree for one sample drawn from this quantile pyramid process, $x$-axis indicates the value of $Q(j/2^m)$.  In figure (b), the lines indicate the intervals from which the quantile values were sampled. Figure (c) shows the interpolated quantile function, the different curves correspond to different samples of the quantile function.}
    \label{fig:pyramid}
\end{figure}

The pyramid quantile process that defines a random probability measure on $[0,1]$ is constructed as follows. 
Let $Q(\tau)$ be the associated random quantile function, with $Q(0)=0$ and $Q(1)=1$. 
At level $m=1$ of the construction the median $Q(1/2)$ is randomly generated over $(0,1)$ according to a given distribution. 
At level $m=2$ of the construction the quartile $Q(1/4)$ is sampled on the interval $(0,Q(1/2))$ and  $Q(3/4)$ is sampled over $(Q(1/2),1)$.   
The process is continued at the following levels $m$, where the quantiles $Q(j/2^m), j=1,3,\ldots, 2^m-1$, are generated conditionally on the quantiles previously sampled.
Figure \ref{fig:pyramid}(a) demonstrates one sample drawn from this quantile pyramid process for $m=1,2,3$, where the value of $Q(j/2^m)$ is indicated on the $x$-axis, and Figure \ref{fig:pyramid}(b) shows the intervals from which successive quantiles at different levels were sampled. 

Specifically, quantiles at level $m$ are generated after those at level $m-1$ according to
\begin{equation}
Q\left(j/2^m\right)=Q\left( (j-1)/2^m\right)(1-V_{mj})+Q\left( (j+1)/2^m\right)V_{mj} \, ,
\label{eq:Qm}
\end{equation}
where $Q\left(j/2^m\right)$ is the new quantile defined at level $m$ and where  $Q\left( (j-1)/2^m\right)$ and $Q\left( (j+1)/2^m\right)$ are its closest ancestors. The independent variable at work at  each level $m$, $V_{mj}$, is a random variable on the unit interval. A \textcolor{black}{natural choice  is to use  $V_{mj}$'s that are Beta distributed,  see \shortciteN{hjortw09} for other possibilities. 
As $m$ tends to infinity the random quantile $Q(\tau)$ is defined for all $\tau$ in $(0,1)$.} 
Notably, the behaviour of this quantile pyramid process depends on these variables. For instance, if at each level we impose that $\mathbb{E}(V_{mj})=0.5$, then we have $\mathbb{E}(Q(\tau)) = \tau$ for all $\tau$ in $(0,1)$ and the quantile process is centred at the uniform quantile function. Theoretical results that concern  $Q(\tau)$ can be found in \shortciteN{hjortw09}. They describe for example 
 relatively mild conditions involving decreasing variances of $V_{mj}$ for growing $m$ that ensure {\it a.s.} the existence of an absolutely continuous $Q(\tau)$.

In practice, to allow a Bayesian inference on the random distribution, the process is stopped at a finite level $M$ and a linear interpolation on the set of quantiles $Q(j/2^M)$, $j=0,1,...,2^M$, completes the process. 
Figure \ref{fig:pyramid}(c) demonstrates three random samples of the piecewise linear quantile functions obtained from the described procedure. 
The  density function corresponding to this linearly interpolated quantile function is piecewise constant, so there is a well defined likelihood function for this random type histogram model. 
Due to the tree nature of the quantile process, the simultaneous density of the $2^M-1$ quantiles can be written as
\begin{equation}
\begin{array}{ll}
\pi\left(Q\left(\frac{1}{2}\right), Q\left(\frac{1}{4}\right), Q\left(\frac{3}{4}\right), \hspace{-10pt} \right.&\left. \ldots, Q\left(\frac{2^{M-1}}{2^M}\right) \right)  \\
&= \underset{m=1}{\overset{M}{\prod}}
\left\{   \underset{j=1,3,\ldots,2^{m-1}} {\prod} \pi_{mj} \left( Q\left(\frac{j}{2^m}\right) \mid Q\left(\frac{j-1}{2^m}\right), Q\left(\frac{j+1}{2^m}  \right) \right) \right\} \; ,
\end{array}
\label{eq:prior}
\end{equation}
where the densities $\pi_{mj}$ can be derived from Equation \ref{eq:Qm}, based on the density of $V_{mj}$, through a simple transform of variables.

\section{Regression modelling with quantile pyramids} \label{sec:regpyramid}

Here we introduce the use of quantile pyramids in the linear  regression setting. We consider the general case  when several conditional quantiles are of interest, say $Q_{\tau}(Y|{\bf x})$ at quantile levels $ \tau=\tau_1,\tau_2,\ldots, \tau_T$ with $\tau_1<\tau_2<\ldots<\tau_T$.  The covariate ${\bf x}=(x_1,...,x_P)$ belongs to a given bounded subset ${\cal X}$ of $\mathbb{R}^P$. In practice   ${\cal X}$ can be taken as the convex hull of the $N$ observed data points ${\bf x}_i$, $i=1,...,N$.  

\subsection{Model formulation}
The starting point for the model formulation is the simple fact that a hyperplane in $\mathbb{R}^{P+1}$ is determined by the values of $P+1$ of its points.  
Let ${\bf x}^0, {\bf x}^1,\ldots,{\bf x}^P$ denote any $P+1$ locations  with corresponding 
$\tau$th conditional quantile denoted by $Q_{\tau}^p, p=0,\ldots,P.$ 
Without loss of generality let ${\bf x}^0=(0,...,0)$, ${\bf x}^1=(1,0,...,0)$, ${\bf x}^2=(0,1,0,...,0)$, $\ldots$, ${\bf x}^P=(0,...,0,1)$. 
The linear quantile regression model for the $\tau$th conditional quantile $Q_{Y}(\tau|{\bf x})$ can be described by the hyperplane passing through these $P+1$ points
\begin{equation}
\begin{array}{ll}
Q_{Y}(\tau|{\bf x}) &= Q^0_{\tau} +\sum_{p=1}^P (Q^p_{\tau}- Q^0_{\tau}) x_p\\
&\equiv \beta_0(\tau) +\sum_{p=1}^P \beta_p(\tau)x_p \; ,
\end{array} 
\label{eq:regressionp}
\end{equation}
where  $\beta_0(\tau)$ and $\beta_p(\tau)$ denote the regression coefficients at $\tau=\tau_1,\tau_2,\ldots, \tau_T$. 
For other other choices of locations ${\bf x}^0,...,{\bf x}^P$,  Equation \ref{eq:regressionp} which is simply the equation of a plane passing through these points has to be modified. In short, the  proposed model for simultaneous linear quantile regression uses $P+1$  independent finite pyramid quantile processes for the quantile functions $Q^p_{\tau}$. 
%The finite processes described in Section \ref{sec:pyramid} appear limited when considering some arbitrary quantile levels $\tau_1<\tau_2<\ldots<\tau_T$ and a real valued response variable not restricted to $[0,1]$. 
Before proceeding to describe the likelihood, we first present some  extensions of these processes that are important in the quantile regression context.
%We will next present some important extensions of the quantile pyramid process presented in Section \ref{sec:pyramid}, before proceeding to describe the likelihood.

\subsection{Oblique quantile pyramid} \label{sec:oblique}
The quantile pyramid  described in Section \ref{sec:pyramid} uses a dyadic partitioning of the probability interval $[0,1]$. In this setting, the induced quantile levels are all fixed and equally spaced. However, in practice, we may be interested in quantiles at specific levels $\tau$.
% over an irregular grid. 

In these circumstances, the quantile level of a child node of the pyramid tree is usually no longer located in the middle point of the quantile levels of its closest ancestors.  We call this general setting oblique quantile pyramid, as opposed to the regular pyramid previously described. To keep the process centred on the Uniform distribution, we now choose $E(V_{mj})$ to reflect this unequal split using the relative distance from the child quantile level $\tau_{mj}$ to its closest ancestors,
\begin{equation}
E(V_{mj}) =  \frac{\tau_{mj} - \tau_{mj}^L} {\tau_{mj}^R - \tau_{mj}^L} \, ,
\label{eq:EVgen}
\end{equation}
where $\tau_{mj}^L$ and $\tau_{mj}^R$ denote its left and right nearest ancestors' quantile levels, respectively. From Equations \ref{eq:Qm} and \ref{eq:EVgen}, it is easy to see that $E(Q(\tau))=\tau$, i.e. under this construction the oblique quantile pyramid is also centred on the Uniform distribution.

The oblique pyramid is constructed via the following procedure. For a sequence of quantiles $Q(\tau_t), t=1,\ldots,T,$ the first level of the pyramid at $m=1$ generates the quantile whose level is halfway into the set of given quantile levels, we will call it the middle quantile level (not to be confounded with the classic median quantile). If $T$ is odd, this is $Q(\tau_{[T/2]+1})$, and given that $V_{11} \sim Beta(\alpha_{11}, \beta_{11})$, we set $\alpha_{11}, \beta_{11}$ such that $E(V_{11}) =\tau_{[T/2]+1}$, as $\tau_{11}^L=0$ and $\tau_{11}^R=1$ per construction. For the next level $m=2$, we proceed by getting the middle quantile levels from the left and right of $Q(\tau_{[T/2]+1})$ to be the next nodes, and choose the corresponding $\alpha's, \beta's$ to satisfy Equation \ref{eq:EVgen}. The process is then continued until all quantiles in the sequence $Q(\tau_t), t=1,\ldots,T,$ have been specified. For identification purposes, if we have an even number of quantile levels, we define the middle value to be the smallest of the two middle quantile levels.

In addition, we choose to have the parameters $\alpha_{mj}, \beta_{mj}$ increasing with the pyramid level $m$, which reduces the prior variance for growing $m$. Throughout this paper, we choose $\alpha_{mj}=2m$ and $\beta_{mj}=\alpha_{mj}*(1-E(V_{mj})/E(V_{mj})$, if $E(V_{mj}) < 0.5$, where $E(V_{mj})$ is calculated using Equation \ref{eq:EVgen}. Otherwise, considering the symmetric nature of the Beta distribution, if $E(V_{mj}) \geq 0.5$, we take $\beta_{mj}=2m$ and $\alpha_{mj}=\beta_{mj}*E(V_{mj})/(1-E(V_{mj}))$. From our experience, this prior is not very informative and gives a good mixing in Markov chain Monte Carlo (MCMC) posterior simulations.

\subsection{Centring the prior} \label{sec:centring}

Using random quantile functions $Q_{\tau}^p,p=0, 1, \ldots, P,$ for the linear model in Equation \ref{eq:regressionp} defines a prior over the quantile planes. This prior should reflect the prior knowledge with respect to the response $Y$.
%must be adequate with the range of values of the response $Y$ and reflect the prior knowledge.}
%When the response variable is not restricted to $[0,1]$ we need to use appropriate quantile processes $Q_{\tau}^p,p=0, 1, \ldots, P$ for the linear model in Equation \ref{eq:regressionp}.}
%For data arising on the reals, the quantile processes $Q_{\tau}^p,p=0, 1, \ldots, P,$ used to construct our linear model in Equation \ref{eq:regressionp} also need to be defined on the reals. 
The pyramid quantile building process described in Section \ref{sec:oblique} is centred on the Uniform distribution  on $[0,1]$.
Let   $Q_{\tau}^{p,unif}$, $p=0, 1, \ldots, P,$ be $P+1$ independent replications of this process. 
In order to use the pyramid quantiles in   Equation \ref{eq:regressionp}, for data $Y$ arising from the reals, we can centre 
 each $Q_{\tau}^p$ process on the quantile function of a Normal distribution ${\cal N}(\mu^p,(\sigma^p)^2)$, $p=0,...,P$, { via} a simple transformation suggested in \shortciteN{hjortw09},
%we could centre the quantile function on the quantiles of $N(\mu, \sigma^2)$ via the transformation 
\begin{equation}
Q_{\tau}^p = \mu^p + \sigma^p\Phi^{-1}(Q^{p,unif}_{\tau}) \, ,
\label{eq:Nprior}
\end{equation}
where $\Phi^{-1}$ denotes the quantile function of the standard normal distribution,  for some mean parameters $\mu^p$ and standard deviation parameters $\sigma^p$. In this case, for each $\tau$ in $(0,1)$,  the median of the random quantile $Q_{\tau}^p$ is  the $\tau$th quantile of a Normal  distribution ${\cal N}(\mu^p,(\sigma^p)^2)$. 
%Here, the median of the random quantile $Q_{\tau}^p$ is equal to the quantile function of a Normal$(\mu,\sigma^2)$, see \shortciteN{hjortw09}. 
 %, and $Q^{p,unif}$ denotes the pyramid quantiles $Q^p_{\tau}$ of Section \ref{sec:pyramid} % 
More generally one can centre the prior on different distributions other than the Normal, depending on the specific prior knowledge available for the particular application at hand, by setting $Q_{\tau}^p=Q_{null}(Q_{\tau}^{p, unif})$ for some arbitrary quantile function $Q_{null}$. Centring the prior on appropriate distributions can be particularly useful for estimating extreme quantiles, as data is scarce at the tails and the pyramid prior is more informative in the tails. However, it is our experience that, for non-extreme quantiles, results are not very sensitive to the default choice of the Normal distribution. For the clarity of exposition,  we use a prior of the form of Equation \ref{eq:Nprior} for the pivotal quantile pyramids $Q_{\tau}^p, p=0,\ldots,P,$ to describe our methodology.

\begin{figure}[!ht]
    \centering
    {\includegraphics[width=8cm, angle=0]{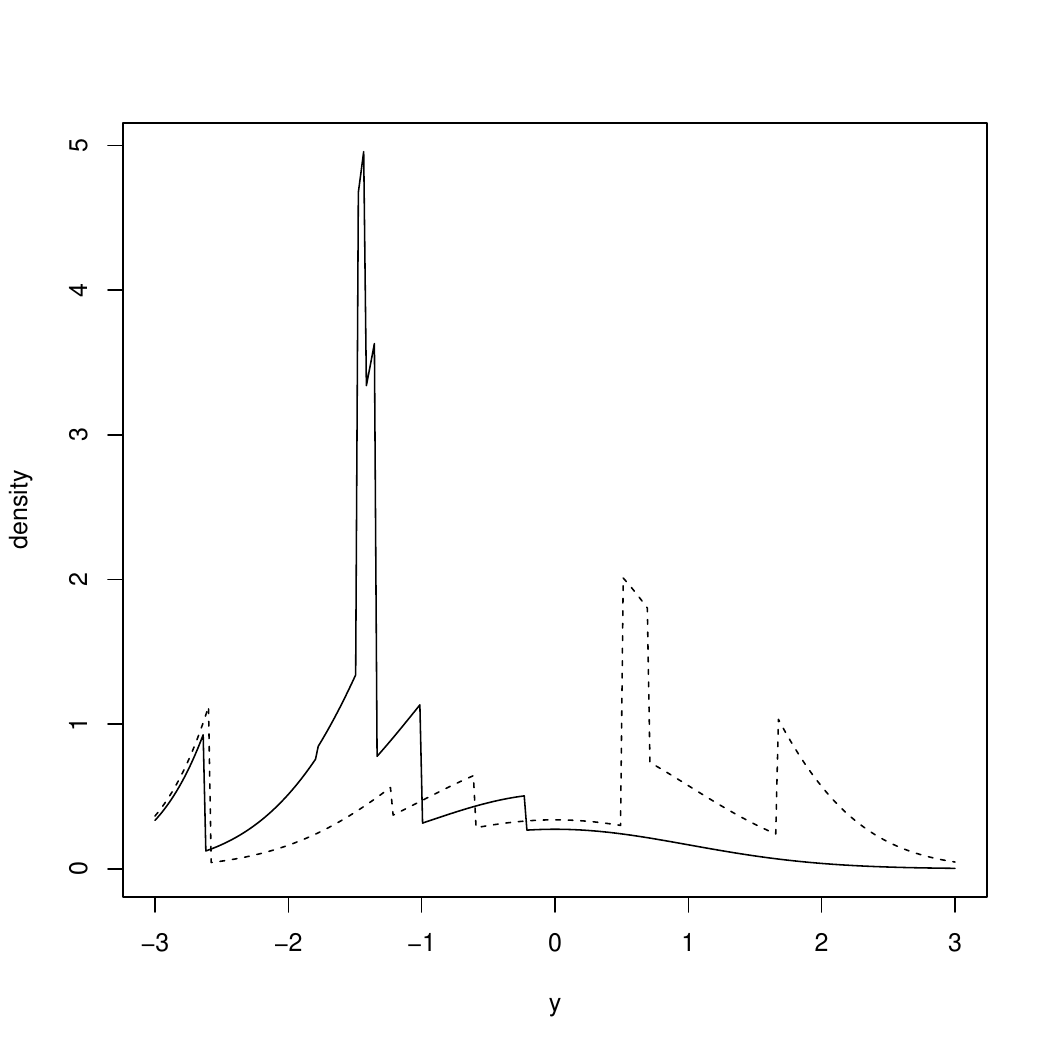}}
    {\includegraphics[width=8cm, angle=0]{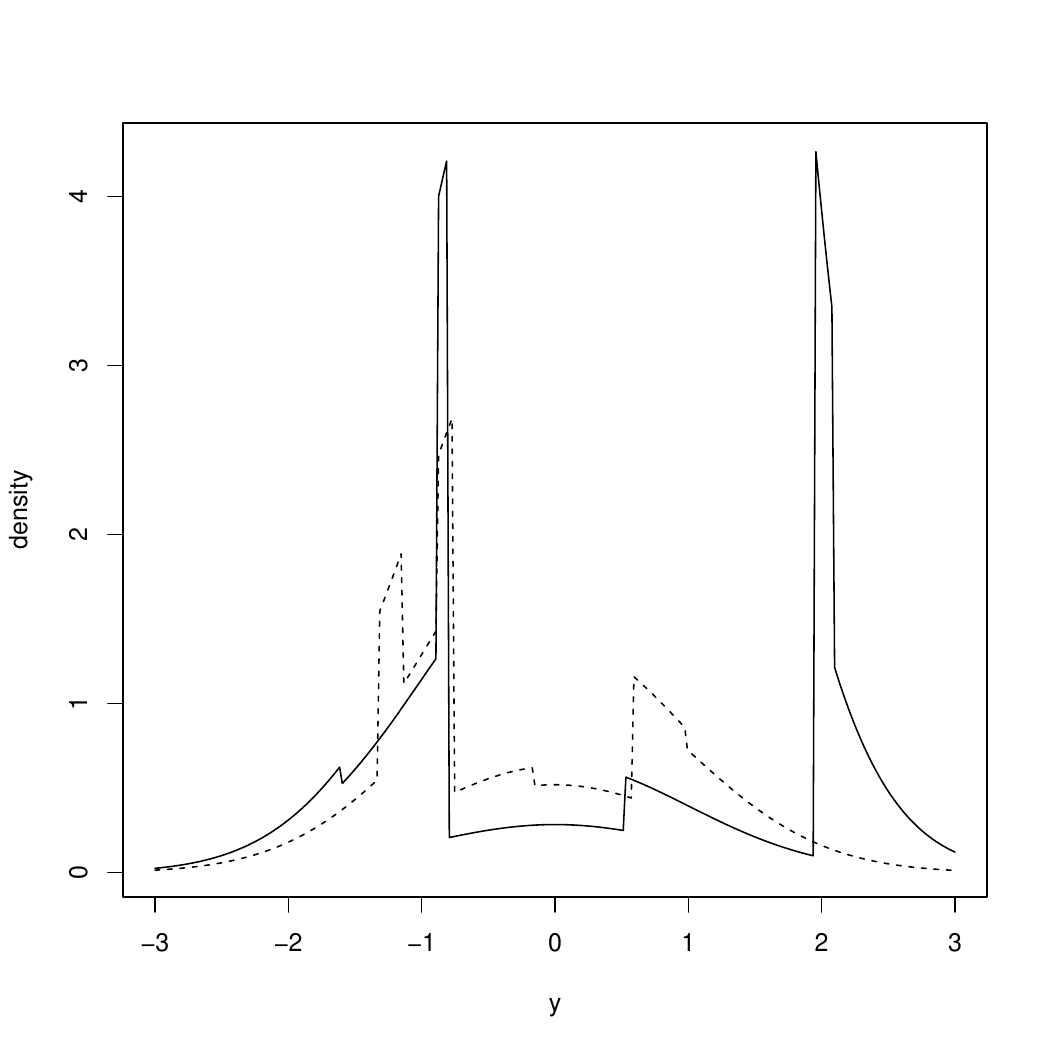}}
    \caption{Examples of simulated densities from a finite pyramid process centred on the standard Normal distribution, with $M=3$ ($\tau=0.125, 0.25, \ldots, 0.875$) and $V_{mj} \sim Beta(a,a)$, for $a=1$ (left) and $a=10$ (right).}
    \label{fig:lik}
\end{figure}

In the finite quantile pyramid context   a random density for $Q_{\tau}^p$ can be derived, which is piecewise scaled Normal distribution between the  quantiles $Q_{\tau_1}^p,\ldots,Q_{\tau_T}^p$. 
%This is a random histogram type model similar to those seen for P\'{o}ya trees. 
This density is obtained by using a simple change of variable on the piecewise constant density function corresponding to $Q^{p,unif}_{\tau}$.
Figure \ref{fig:lik} illustrates some samples of this quantile process, highlighting the piecewise Normal density feature. The examples were simulated from a pyramid process centred on the standard Normal distribution, with $M=3$, $\tau=0.125, 0.25, \ldots, 0.875$ and $V_{mj} \sim Beta(a,a)$, for $a=1 \text{ and } 10$.

\subsection{Likelihood and posterior}

Equation \ref{eq:regressionp} gives the desired quantiles of the conditional distribution of $Y$ given $X={\bf x}$, with cdf $F(y|{\bf x})$. 
When  priors of the form \ref{eq:Nprior} are used, we need to {define} the likelihood function.
The chosen option here is to consider that the density $f(y|{\bf x})$ of the conditional distribution is  piecewise Normal 
\begin{equation}
f(y |{\bf x})= \sum_{t=1}^{T} {(\tau_t-\tau_{t-1})\frac{\phi(y; \mu_{{\bf x}},\sigma_{{\bf x}}^2)}{ \Phi \left (\frac{Q_{Y}(\tau_t|{\bf x})-\mu_{{\bf x}}}{\sigma_{{\bf x}}}\right )-\Phi \left(\frac{Q_{Y}(\tau_{t-1}|{\bf x})-\mu_{{\bf x}}}{\sigma_{{\bf x}}}\right)}I_{(Q_{Y}(\tau_{t-1}|{\bf x}), Q_{Y}(\tau_{t}|{\bf x})]}(y)},
\label{eqnLik}
\end{equation}
where $I_{(q_1, q_2]}(y)$ is 1 if $y \in (q_1, q_2 ]$ and zero otherwise, where $\phi(\cdot; \mu,\sigma^2)$ denotes the density function of a Normal distribution ${\cal N}(\mu,\sigma^2)$ and where the parameters $\mu_{{\bf x}}$ and $\sigma_{{\bf x}}$ change linearly in ${\bf x}$
\begin{equation}
\mu_{\bf x}= \left( 1 -\sum_{p=1}^P x_p \right) \mu^0+ \sum_{p=1}^P x_p \mu^p \, , \quad\quad 
\sigma_{\bf x}= \left(  1 -\sum_{p=1}^P x_p \right) \sigma^0+ \sum_{p=1}^P x_p \sigma^p \, .
\label{eq:planemu}
\end{equation}
This formulation implies that the priors on all conditional distributions are centred on the Normal distribution 
 and Equation \ref{eq:planemu} specifies that the quantiles of these centring distributions  change linearly in the covariates. {This 
 additional assumption on the form of the prior is quite natural in the linear quantile setting and not overly restrictive. }
Equation \ref{eqnLik} can be obtained by extending the random histogram-type likelihood corresponding to the finite quantile pyramid centred on the uniform distribution as in \shortciteN{hjortw09} and applying the relevant transformation of the Equation \ref{eq:Nprior}.

Note that {a more general} approach {which does not require the assumption of Equation \ref{eq:planemu}, is to specify the likelihood function by working directly} 
% is to specify that all quantile levels vary linearly with x, ie. assume Equation \ref{eq:regressionp} for all $\tau \in [0,1]$, and obtain the density of the conditional distribution considering that $f(y|{\bf x})=\frac{1}{Q'_{F(y_i|{\bf x})}(Y|{\bf x})}$.}
%Note that an alternative approach is to 
with the density of the conditional distribution $f(y|{\bf x})=\frac{1}{q_{\bf x}(F(y_i|{\bf x}))}$ where $q_{\bf x}$ denotes the quantile density  at ${\bf x}$, {\it i.e.} the derivative of $Q_{Y}(\tau|{\bf x})$ with respect to $\tau$. 
%This derivative can be deduced from Equation \ref{eq:regressionp}. 
Nevertheless, numerical search over a fine grid is required for the evaluation of the density at each data observation, which
increases both the numerical error and computational burden. We therefore choose to work with Equation  \ref{eq:planemu} in this article.

%the value $F(y|{\bf x})$ can only be approximated and the involved numerical cost to approximate $f(y_i|{\bf x}_i)$ for each observation $i=1,...,N$ can be prohibitive.
%\textcolor{red}{Furthermore, in practice only a finite number of quantiles are needed, and the proposed likelihood \eqref{eqnLik} is a direct consequence of the piecewise Normal prior \eqref{eq:Nprior}.}
%Furthermore the implied conditional density of this approach is not piecewise Normal at each ${\bf x}$.

The posterior distribution for the finite number of quantile levels $\tau_1,\ldots, \tau_T$ can be obtained for the quantiles $\mathbf{Q}^p=\{Q^p_{\tau_1}, \ldots, Q^p_{\tau_T}\}, p=0,\ldots,P,$ and the associated parameters ${\boldsymbol \mu} = \{ \mu^0,\ldots,\mu^P\}, {\boldsymbol \sigma}=\{\sigma^0,\ldots, \sigma^P\}$, via the usual Bayes theorem 
\begin{equation}
\pi(\mathbf{Q}^0, \ldots, \mathbf{Q}^P, {\boldsymbol \mu}, {\boldsymbol \sigma}|y_1,\ldots, y_N) \propto \prod_{i=1}^N f(y_i|{\bf x}_i) \times \prod_{p=0}^P \pi(\mathbf{Q}^p |\mu^p, \sigma^p)  \times \pi({\boldsymbol \mu}) \times \pi({\boldsymbol \sigma}) \, ,
\label{eq:post}
\end{equation}
where $f(y_i|{\bf x}_i)$ is given in Equation \ref{eqnLik}. The distributions $\pi({\boldsymbol \mu})$ and $ \pi({\boldsymbol \sigma})$ are hyperpriors for the parameters of the Normal distributions. Throughout the paper these hyperpriors are set to $N(0,20)$ and $\text{Gamma}(0.001,0.001)$, for $\mu^p$ and $\sigma^p$ respectively. In addition, using Equations \ref{eq:prior} and \ref{eq:Nprior}, the pivotal pyramid prior distributions $\pi(\mathbf{Q}^p |\mu^p, \sigma^p),$ $p=0,\ldots, P,$ are
\begin{subequations}
\begin{align*}
\pi(\mathbf{Q}^p|\mu^p, \sigma^p) &=  \prod_{m,j}^{} { \pi_{mj} \left( Q^p_{\tau_{mj}} | Q^p_{\tau_{mj}^L}, Q^p_{\tau_{mj}^R} \right) } \\
&= \prod_{m,j}^{} \left\{ g\left(\frac{\Phi \left(\frac{Q^p_{\tau_{mj}} - \mu^p} {\sigma^p} \right)-\Phi \left(\frac{Q^p_{\tau_{mj}^L} - \mu^p} {\sigma^p}\right)}{\Phi \left(\frac{Q^p_{\tau_{mj}^R} - \mu^p} {\sigma^p}\right)-\Phi \left(\frac{Q^p_{\tau_{mj}^L} - \mu^p} {\sigma^p}\right)}\right) \times \frac{\phi \left(\frac{Q^p_{\tau_{mj}} - \mu^p} {\sigma^p}\right)}{\Phi \left(\frac{Q^p_{\tau_{mj}^R} - \mu^p} {\sigma^p}\right)-\Phi \left(\frac{Q^p_{\tau_{mj}^L} - \mu^p} {\sigma^p}\right)}   \right\} \; ,
 \end{align*}
\end{subequations}
where $g(\cdot)$ denotes the density of the $V_{mj}$ variables, throughout the paper $V_{mj} \sim Beta(\alpha_{mj}, \beta_{mj})$, and $\phi$ the standard Normal density.

\subsection{Non-crossing constraints}\label{mcmc}

The linear model proposed in this paper ensures that the simultaneously fitted quantile planes in Equation \ref{eq:regressionp}  do not cross on the convex hull of the $P+1$ pivotal locations ${\bf x}^0,{\bf x}^1,\cdots,{\bf x}^P$.
For the single covariate problem, choosing pivotal locations ${\bf x}^0$ and ${\bf x}^1$  to be the minimum and maximum value of ${\cal X}$ is sufficient to ensure non-crossing.  
However, for $P>1$,  some caution is needed with crossings. If ${\cal X}$ corresponds to the convex hull of the observed data points and the  pivotal quantiles are placed at $P+1$ well separated vertices, the non-crossing of the planes needs to be verified at any remaining vertices of the convex hull, say at some points denoted ${\bf x}^e$'s. 
Note that working with any convex sets larger than the minimum convex set enclosing the data will also ensure non-crossing, but convex sets that are too large puts unnecessary constraints on the regression model, forcing the regression planes to be parallel.

A naive option to ensure non-crossing is to check for crossing at the non-pivotal locations of the convex hull and discard the samples that produce crossing planes during the Metropolis-Hastings MCMC sampling procedure.
However, for moderate numbers of covariates, this approach is very inefficient as the crossing will most likely be frequent. In fact, a better solution is to adjust the MCMC proposal distribution so that it proposes only in the non-crossing region. To accomplish that, we will adopt Uniform proposals $U(l_\tau^p,u_\tau^p)$ for $Q^p_\tau$, and choose the lower and upper bounds ($l_\tau^p$, $u_\tau^p$) while ensuring that the corresponding quantiles at the non-pivotal locations do not cross, \shortciteN{fengch2015} used a similar approach working with the entire dataset.

More specifically, for each extra location ${\bf x}^{e}$ on the vertices of the convex hull, the bounds can be easily found by solving $Q_{\tau}^e = Q_{\tau-1}^e$ and $Q_{\tau}^e = Q_{\tau+1}^e$ for $Q^p_\tau$ based on the hyperplane equation. For example, for the hyperplane described in Equation \ref{eq:regressionp}, when the $p$th component of ${\bf x}^e$ is greater than zero, i.e. $x_p^e >0$, we have
\begin{subequations}
\begin{align*}
 l_\tau^p|{\bf x}^e &=  Q_\tau^0 + \left( Q_{\tau-1}^e -  Q_{\tau}^0  - \sum_{j \neq p} \left( Q_\tau^j - Q_\tau^0 \right) x_j^e \right) / x_p^e \; , \\
 u_\tau^p|{\bf x}^e &=  Q_\tau^0 + \left( Q_{\tau+1}^e -  Q_{\tau}^0  - \sum_{j \neq p} \left( Q_\tau^j - Q_\tau^0 \right) x_j^e \right) / x_p^e \; ,
\end{align*}
\end{subequations}
where $l_\tau^p|{\bf x}^e$ and $u_\tau^p|{\bf x}^e$ are $Q^p_\tau$ lower and upper bounds, respectively, based on the crossing restrictions at ${\bf x}^e$. Similarly, if $x_p^e<0$, the above lower bound becomes the upper bound and vice-versa. Therefore, to take into account all non-pivotal vertices' constraints, we choose $l_\tau^p = \text{max}\{l_\tau^p|{\bf x}^e\}$ and $u_\tau^p = \text{min}\{u_\tau^p|{\bf x}^e\}$. In this way, non-crossing issues are easily handled. Note that the bounds for each extra location can be found at once through simple matrix operations, not being computationally very expensive.

\subsection{Large pyramidal support and posterior consistency}  \label{sec:consistency}
In this section we first study the support of the proposed prior on the quantile planes provided by infinite quantile pyramids. We then give a posterior consistency  property of the procedure that uses finite quantile pyramids defined upon a level $M_n$ that grows slowly with the sample size $n$.

For a formal treatment of these topics in the regression context we follow  \shortciteN{yang2015} and consider a stochastic design setting where the covariates $X_i$'s are drawn from a pdf $f_X$. 
In order to ensure that the linear model (\ref{eq:regressionp}) is valid we suppose here that the support ${\cal X}$ of $f_X$ is a subset of the convex hull of the pyramid locations ${\bf x}^0,\dots,{\bf x}^P$. By using infinite quantile pyramids $Q_{\tau}^0,\dots,Q_{\tau}^P$, we define a prior probability measure $\Pi$ on the set ${\cal F}=\{f(x,y)=f_X(x)f_Y(y|x)\}$ of density functions on ${\cal X}\times \mathbb{R}$.

Let $f^*(x,y)=f_X(x)f_Y^*(y|x)$ be a given density functions on ${\cal X}\times \mathbb{R}$, that later will be considered as the true data generating process. 
Let $d_{KL}(f^*,f)=\int f^* \ln(f^*/f)$ denotes the KL divergence between $f^*$ and $f$. 
By extending  Proposition 3.1 in \shortciteN{hjortw09} to the regression setting we first show that, under some regularity conditions, $f^*$ is in the Kullback-Leibler (KL) support of $\Pi$, {\it i.e.} for any $\epsilon>0$ we have 
$\Pi \left( \{ f:\, d_{KL}(f^*,f)<\epsilon  \}\right)>0$.   

To do this, if for $p=1,\dots,P$ we take $Q_{\tau}^p=Q_{null}^p(Q_{\tau}^{p,unif})$, % where the quantile processes $Q_{\tau}^{p,unif}$ are centered on the uniform distribution, 
we first suppose that the  conditions  ensuring that the processes $Q_{\tau}^{p,unif}$ are a.s. absolutely continuous  are verified, 
and let $q_p^{unif}(\cdot)$ be the corresponding quantile density function. 
For each $p=1,\dots,P,$ let also $q_p^*(\cdot)$ be the quantile density function corresponding to the density $f_Y^*(y|{\bf x}^p)$ 
and let  $q_p^{*unif}(\cdot)$ be the quantile density function corresponding to the quantile function $Q_p^{*unif}(\cdot)=F_{null}^p(Q_p^{*}(\cdot))$. 
%\textcolor{red}{should this be $F_{null}$ not $F^*_{null}$?}
The regularity conditions are simply the conditions (A)-(C) described in \shortciteN{hjortw09} applied at each pyramid location ${\bf x}^0,\dots,{\bf x}^P$ plus a regularity condition on the centring quantile functions $Q_{null}^p$. 
More precisely we consider the following conditions\begin{description}
\item[(A)]
for any $\epsilon>0$ and for all $p=1,\dots,P$ we have $\Pi\left(\{q_p^{unif}:\, \int q_p^{unif} \ln(q_p^{unif}/q_p^{*unif})<\epsilon \}\right)>0$,
\item[(B)]
for all $\delta>0$ and for all $p=1,\dots,P$ there exists an $\epsilon>0$ such that
$$
\int \ln\frac{q_p^*(\tau_{\epsilon}(u))}{q_p^*(u)}du <\delta
$$
for any function $\tau_{\epsilon}(u)$ with values in $(0,1)$ for which $\max_u |\tau_{\epsilon}(u)-u|<\epsilon$,
\item[(C)]
for each $p=1,\dots,P$ the density $f_Y^*(y|{\bf x}^p)$ is bounded by a finite value,
\item[(D)]
for each $p=1,\dots,P$ the quantile function $Q_{null}^p$ is absolutely continuous.
\end{description}
\begin{proposition}
Under the conditions (A)-(D)  the density $f^*$ is in the KL support of $\Pi$.
\label{KLsupport}
\end{proposition}
The proof is given in the Appendix. The smoothness condition (B) and the condition of boundary (C) concern only the density $f^*$. Concerning  (A),   \shortciteN{hjortw09} have shown that this condition is verified by a quantile pyramid $Q_{\tau}^{unif}$ on $[0,1]$ when the $V_{mj}$'s have expectations fixed at 0.5 and  variances decreasing sufficiently fast, more precisely $\sum_{m=1}^{+\infty} \max_j Var(V_{jm})<+\infty$. 
%Under the same hypotheses on the   $V_{mj}$'s the condition  is also verified by a quantile pyramid of the form $Q_{null}(Q_{\tau}^{unif})$ if $Q_{null}$ is absolutely continuous wr.t. the Lebesgue measure on $(0,1)$.
The condition (D) is fulfilled by any  quantile function that admits a derivative and  corresponds to a distribution with a bounded support. 
This is not verified for example in the case when the centring distribution is Gaussian but, in practice,   one can consider  instead a truncated version  on an arbitrarily large interval. 
%A Gaussian quantile function does not fulfill this condition but one can consider instead a truncated version. 
%\textcolor{red}{The gaussian quantile function is defined on (0,,1), so its on bounded support, do you mean that $F_{null}$ has to be on bounded support? }

In practice we use finite quantile pyramids defined until a finite level $M$. 
A common practice is to use a level $M$ that is size dependent, say $M_n$, increasing with $n$. 
In this case, again by extending a result from \shortciteN{hjortw09}, we can establish a strong consistency property, called Hellinger consistency, of the resulting prior $\Pi_{M_n}$. 
%More precisely we consider the Hellinger consistency. 
Let  $f^*(x,y)=f^X(x) f_x^*(y)$ be a density  in the KL support of $\Pi$, the prior constructed with infinite pyramid quantile processes  and let $(X_i,Y_i), \, i=1,\dots,n,\dots$ be independent observations from $f^*$. 
The Hellinger distance $d_H(f^*,f)$ between the densities $f^*$ and $f$ is defined as $d_H^2(f^*,f)=\int \left( \sqrt{f^*}- \sqrt{f} \right)^2$. 
The sequence of posterior distributions $\{\Pi_{M_n}(\cdot|(X_i,Y_i), \, i=1,\dots,n)\}_n$ is said to be Hellinger consistent at $f^*$ if, for every $\epsilon>0$ and for every set
$$
A_{\epsilon}=\{f(x,y): d_H^2(f^*,f)<\epsilon\}
$$
we have $\Pi_{m_n}(A_{\epsilon}|(X_i,Y_i), \, i=1,\dots,n)\rightarrow 1$ a.s.
\begin{proposition}
Under the conditions (A)-(C)  and if $M_n$ is such that $M_n\rightarrow +\infty$ and $2^{M_n}/n\rightarrow0$ then the sequence of posterior distributions $\{\Pi_{M_n}(\cdot|(X_i,Y_i), \, i=1,\dots,n)\}_n$ is  Hellinger consistent at $f^*$.
\label{HLconsistency}
\end{proposition}

The proof for this proposition is given in the appendix.

\section{Simulated examples} \label{sec:sim}
In this section, small sample properties of the pyramid quantile regression estimator (PQR) will be investigated through simulation examples. Also, PQR will be compared with three other approaches:  semiparametric regression model (BSquare) of \shortciteN{reichs13}, Gaussian process method (GPQR) of \shortciteN{yang2015}  and the frequentist constrained estimator (freqQR) of \shortciteN{Bondell2010}. The comparisons will be undertaken in terms of $95\%$ coverage probabilities and the empirical root mean squared error $RMSE(\tau) = \sqrt{ 1/s \sum_{s=1}^{S} { [\beta(\tau) - \hat{\beta}_s(\tau)]^2}}$, based on $S=200$ data sets. Following \shortciteN{reichs13}, we use the simulation designs that are detailed below. 

\begin{description}
 \item [Design 1.] $\beta_0(\tau)=\log[\tau/(1-\tau)]$, $\beta_1(\tau)=2$;
 \item [Design 2.] $\beta_0(\tau)=\text{sign}(0.5-\tau)\log{(1-2 \left|0.5-\tau \right|)}$, $\beta_1(\tau)=2\tau$;
 \item [Design 3.] $\beta_0(\tau)=\Phi^{-1}(\tau)$, $\beta_1(\tau)=2 \min{\{\tau-0.5,0\}}$;
 \item [Design 4.] $\beta_0(\tau)=2\Phi^{-1}(\tau)$, $\beta_1(\tau)=2 \min{\{\tau-0.5,0\}}$, $\beta_2(\tau)=2\tau$, $\beta_3=2$, $\beta_4=1$, $\beta_5=0$;
\end{description}

\begin{figure}[!ht]
    %\captionsetup[subfigure]{labelformat=empty}
    \centering
    \includegraphics[width=5cm,height=8cm,angle=-90]{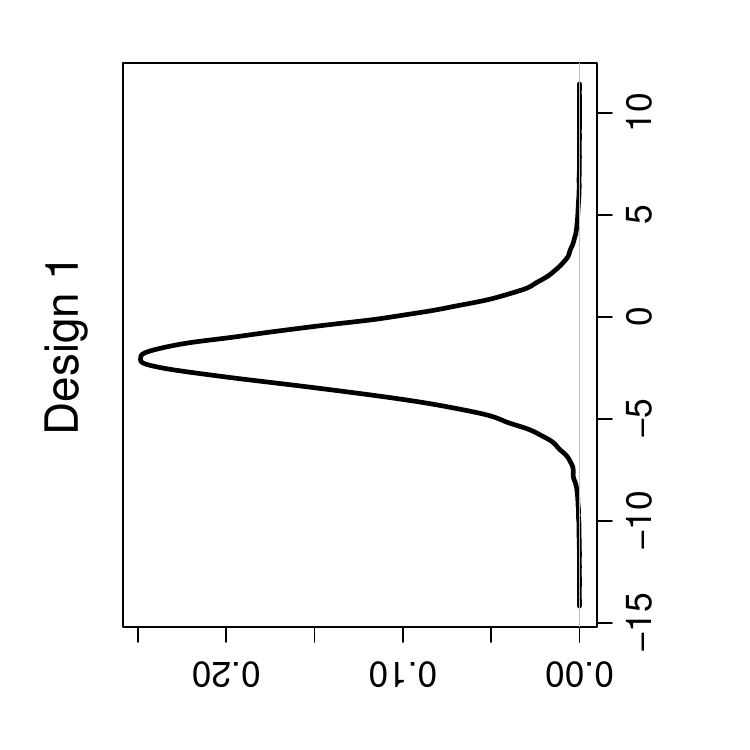}
    \includegraphics[width=5cm,height=8cm,angle=-90]{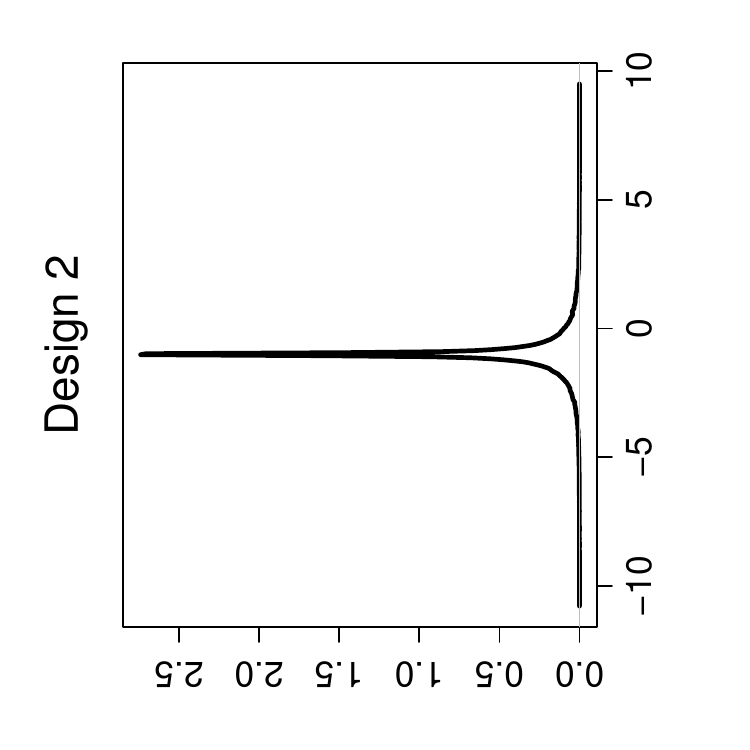} \\
    \includegraphics[width=5cm,height=8cm,angle=-90]{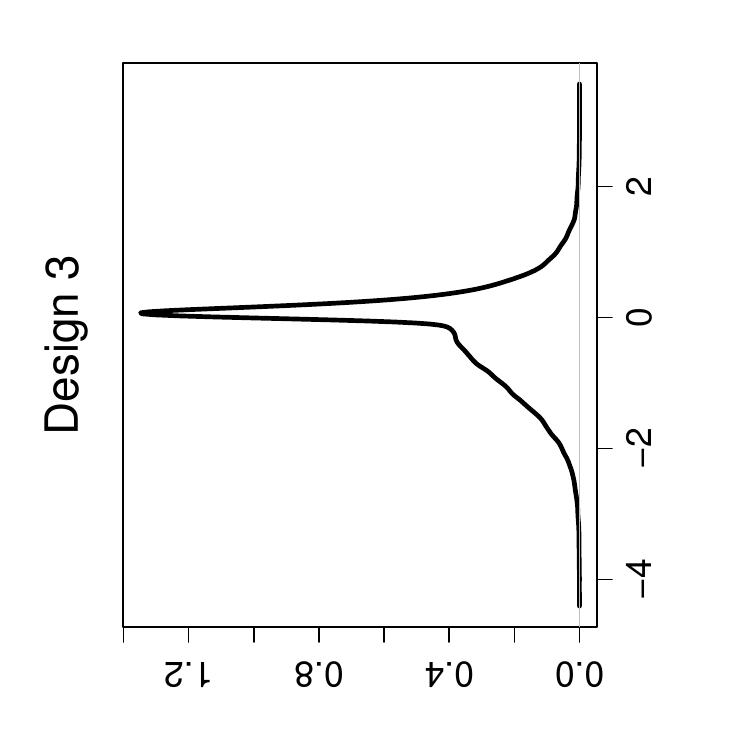}
    \includegraphics[width=5cm,height=8cm,angle=-90]{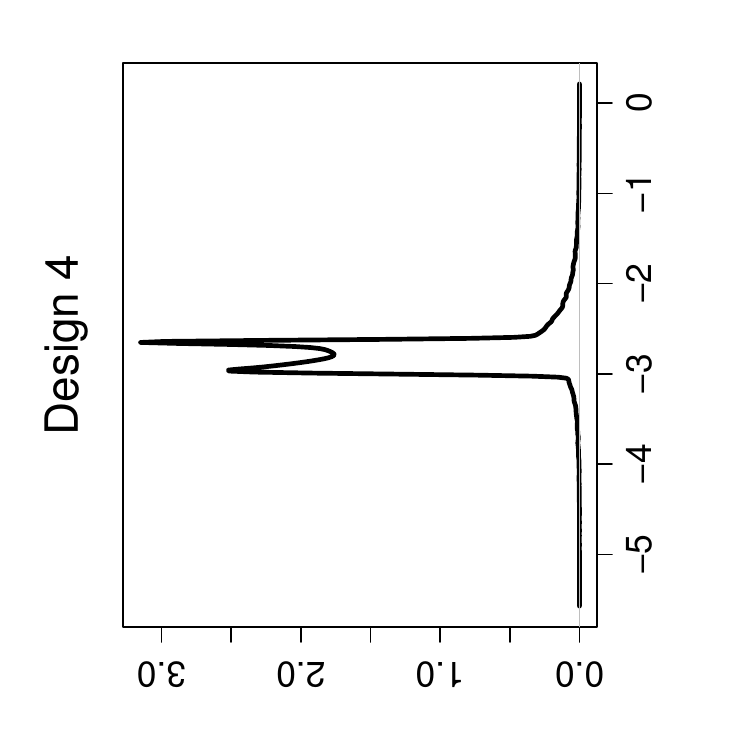}
    \caption{True conditional densities $f(Y|x=-1)$ for the simulation designs.}
    \label{fig:true_den}
\end{figure}

For each design, we simulated some observations $y_i$, $i=1,...,N$, from
$$
y_i = \beta_0(u_i)+ \sum_{j=1}^P x_{ij}\beta_j(u_i), 
$$
where the $j$-th covariate is $x_{ij} \overset{iid}{\sim} \text{Unif}(-1,1)$ and $u_i \overset{iid}{\sim} \text{Unif}(0,1)$. The simulated conditional densities at $x=-1$, $f(Y|x=-1)$, for designs $1$ to $4$ are illustrated in Figure \ref{fig:true_den}.

% \begin{figure}[!ht]
%     %\captionsetup[subfigure]{labelformat=empty}
%     \centering
%     \subfloat[Design 1 \label{dens1}]{ \includegraphics[width=5cm,height=5cm,angle=-90]{./Rfigures/des1}}
%     \subfloat[Design 2 \label{dens2}]{ \includegraphics[width=5cm,height=5cm,angle=-90]{./Rfigures/des2}} \\
%     \subfloat[Design 3 \label{dens3}]{ \includegraphics[width=5cm,height=5cm,angle=-90]{./Rfigures/des3}}
%     \subfloat[Design 4 \label{dens4}]{ \includegraphics[width=5cm,height=5cm,angle=-90]{./Rfigures/des4}}
%     \caption{True conditional distributions of $Y|x=-1$ under Designs 1 to 3.}
%     \label{fig:true_den}
% \end{figure}

For univariate designs $1$ to $3$, we used the datasize $N=100$ and estimated simultaneously the quantile regression lines at quantile levels $\tau = 0.01,0.05,0.10,...,0.95,0.99$. PQR was fitted based on $110.000$ MCMC draws and burn-in of $10.000$. Furthermore, in order to improve MCMC mixing, the pyramid quantiles were reparametrised using the logarithm of the difference between adjacent quantile levels, i.e. $\{ \log{(Q^p(\tau_2) - Q^p(\tau_1))} , \ldots ,$ $\log{(Q^p(\tau_T) - Q^p(\tau_{T-1}))}, \log{(Q^p(\tau_1) + Q^p(\tau_T) + c )}\}$, where a constant $c=2|\text{min}(Y_i)|$ was added to the last term to prevent a negative argument in the logarithm function. Posterior means were taken as point estimates for the $\beta$'s.

BSquare estimator is implemented in BSquare package (\shortciteNP{BSquare}) in R (\shortciteNP{Rmanual}), to fit this model we used the logistic base distribution with $4$ basis functions. GPQR model is also available in R (qrjoint package by \shortciteNP{Rqrjoint}), and it was estimated from $50.000$ MCMC samples, thinning every $10$ samples and discarding the initial $20\%$ of the samples as burn-in. Codes for \shortciteN{Bondell2010} are available from first author's web page.

\begin{figure}[!ht]
    %\captionsetup[subfigure]{labelformat=empty}
    \centering
    \includegraphics[width=8cm,height=7.5cm,angle=-90]{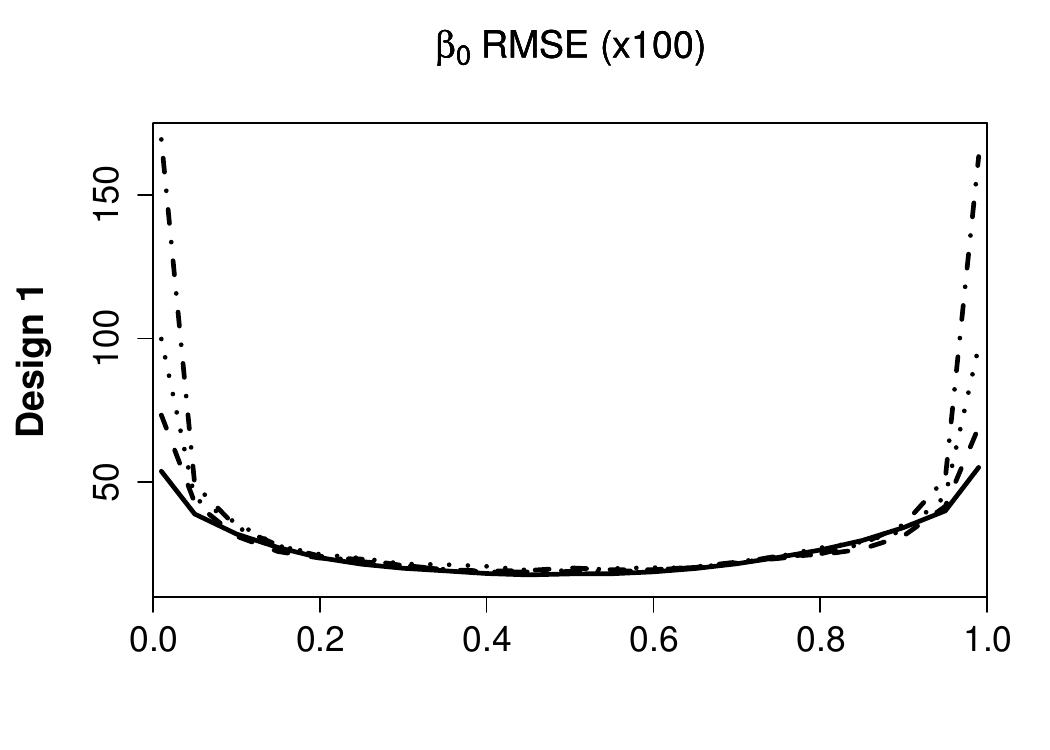}
    \includegraphics[width=8cm,height=7.5cm,angle=-90]{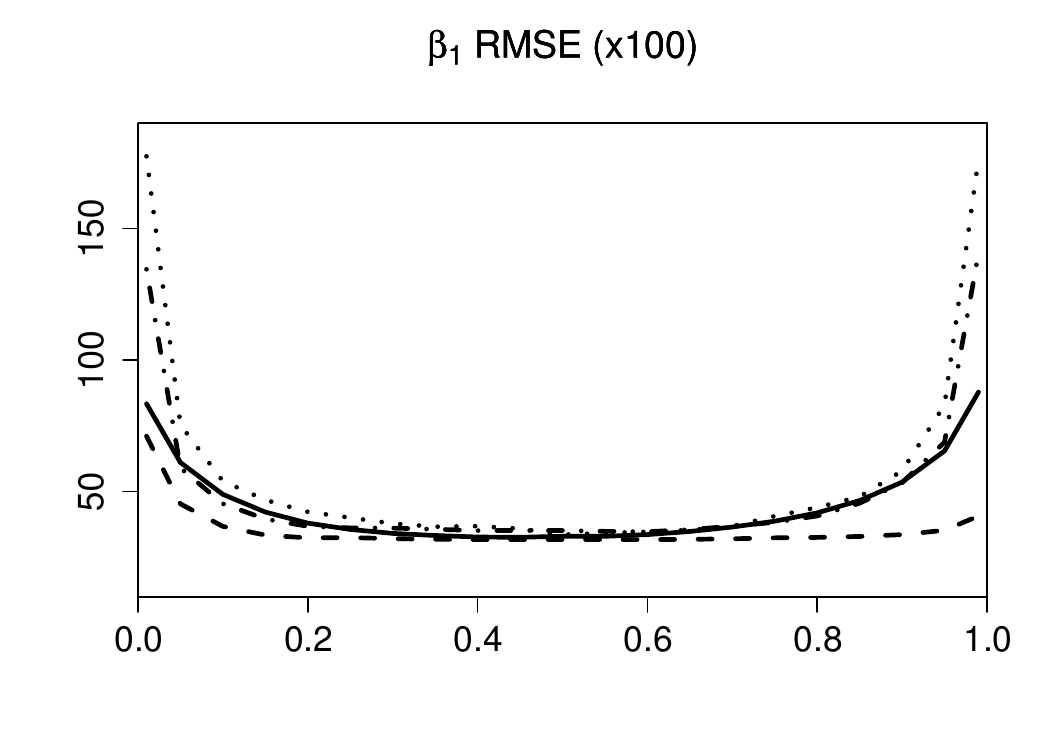} \\[-1.5cm]
    \includegraphics[width=8cm,height=7.5cm,angle=-90]{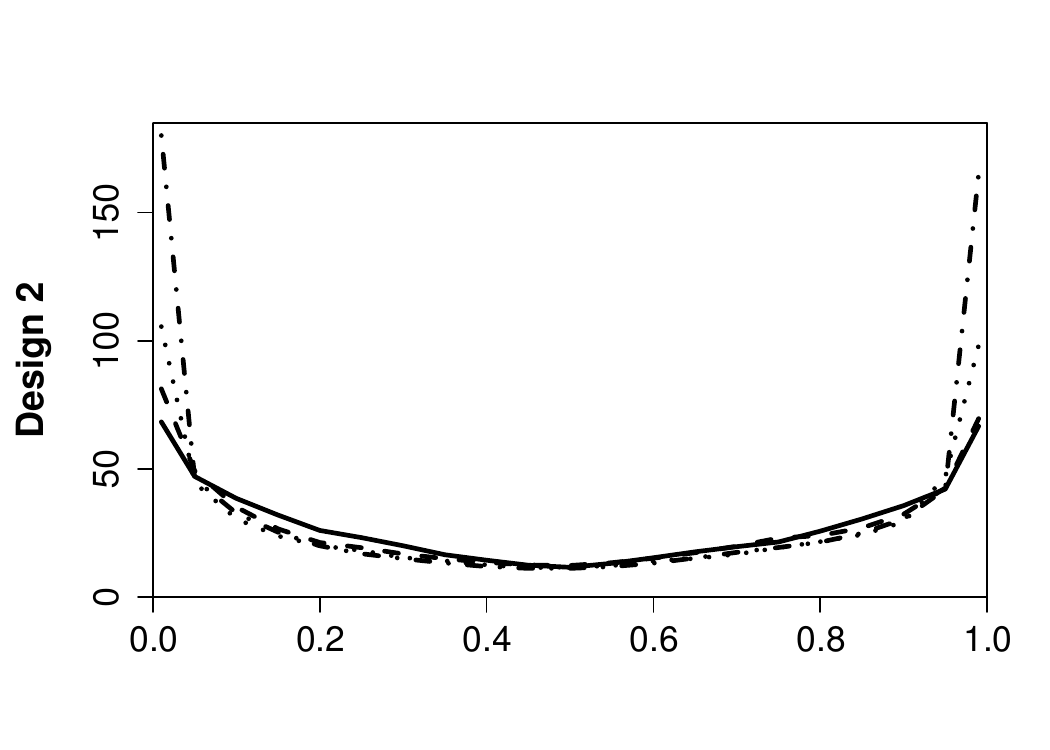}
    \includegraphics[width=8cm,height=7.5cm,angle=-90]{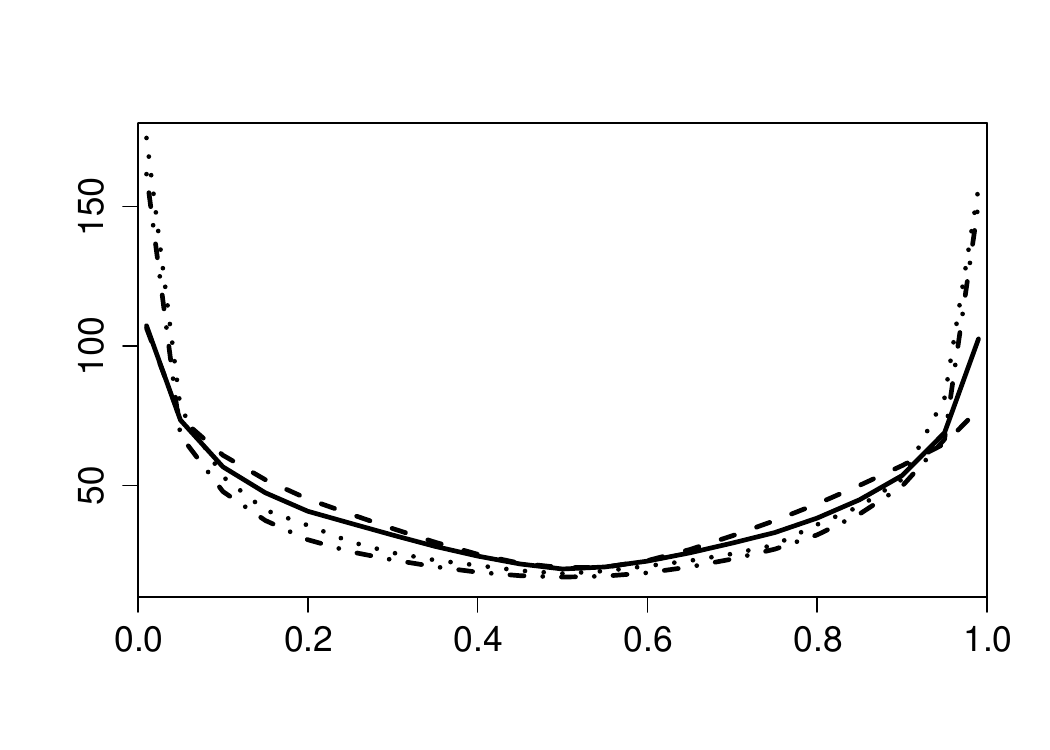} \\[-1.5cm]
    \includegraphics[width=8cm,height=7.5cm,angle=-90]{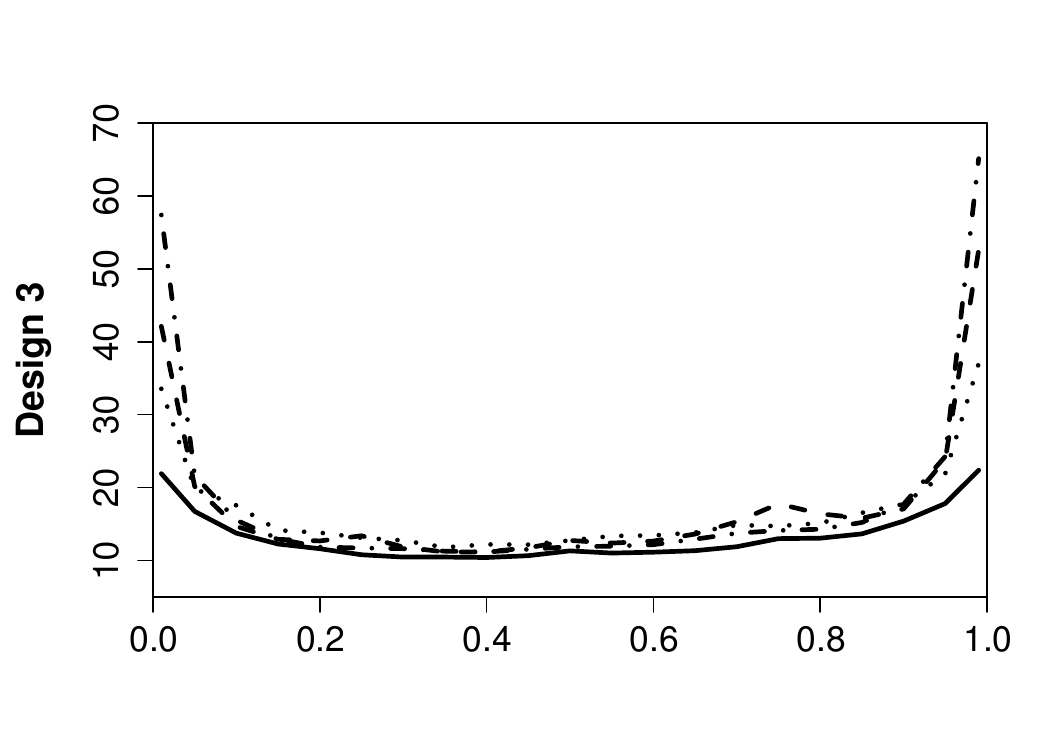}
    \includegraphics[width=8cm,height=7.5cm,angle=-90]{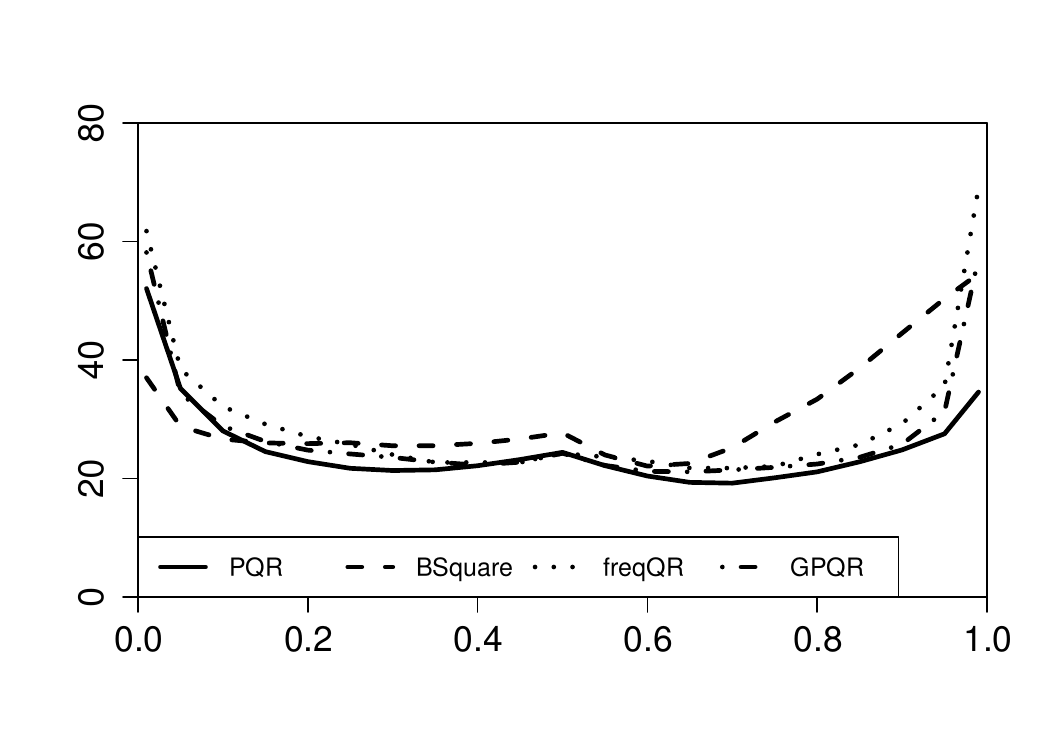}
    \caption{RMSE ($\times 100$) for $\beta_0$ (left) and $\beta_1$ (right)  at $\tau=0.01,0.05,0.1,\dots,0.95, 0.99$.}
    \label{fig:sim_linear_mse}
\end{figure}

Figure \ref{fig:sim_linear_mse} presents RMSE results for the univariate designs. Overall we can see that, for non-extreme quantile levels, all methods perform similarly, with BSquare having the best results for $\beta_1$ from Design $1$ and PQR  having the best results for $\beta_1$ from Design $3$. Data from design $1$ follows BSquare model assumptions, which certainly contributes to its better performance. Design $3$ presents a more challenging quantile function, and the flexibility of the proposed approach is an advantage here.

For extreme quantiles, PQR clearly outperforms the other methods for most cases. Once again the flexibility of the proposed approach contributes to this achievement, as well as the reasonable choice of the quantile process centring distribution. Note that, although the simulated designs are not from a Normal distribution (e.g. see Figure \ref{fig:true_den}), yet this is a reasonable centring choice here. The meaningfulness of quantile parameters in PQR is a great feature of the proposed model, as prior information are easily interpreted and incorporated. As shown here, a rough idea of the true distribution can contribute to improve the estimation of extreme quantiles. 

\begin{figure}[!htbp]
    %\captionsetup[subfigure]{labelformat=empty}
    \centering
    \includegraphics[width=8cm,height=7.5cm,angle=-90]{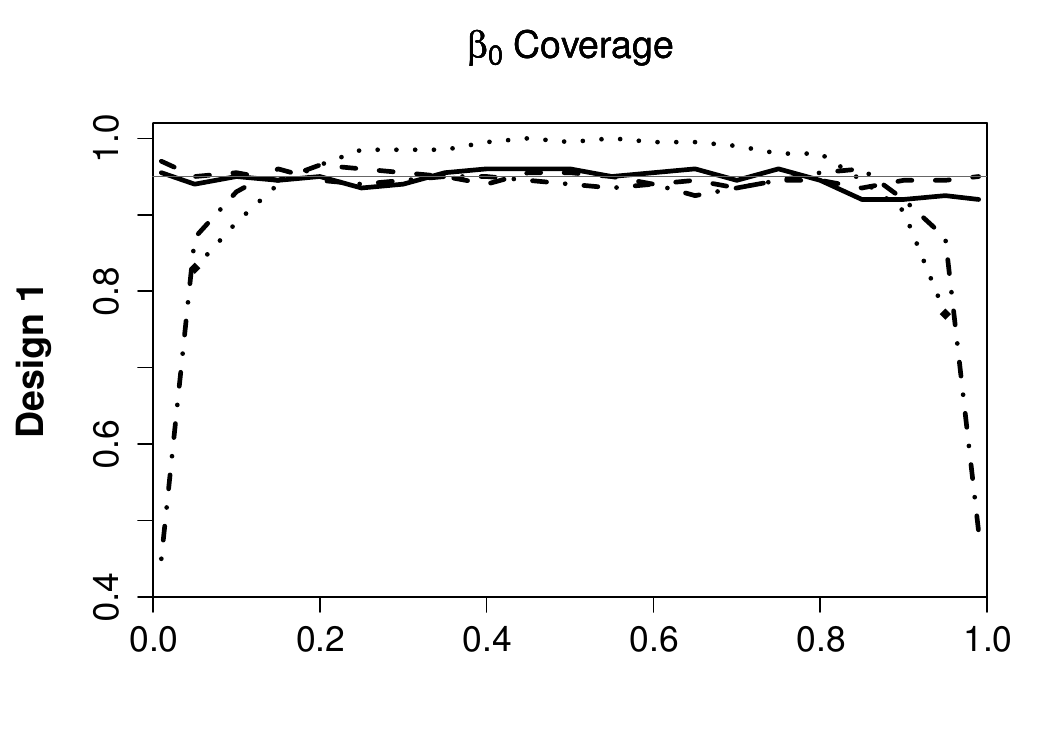}
    \includegraphics[width=8cm,height=7.5cm,angle=-90]{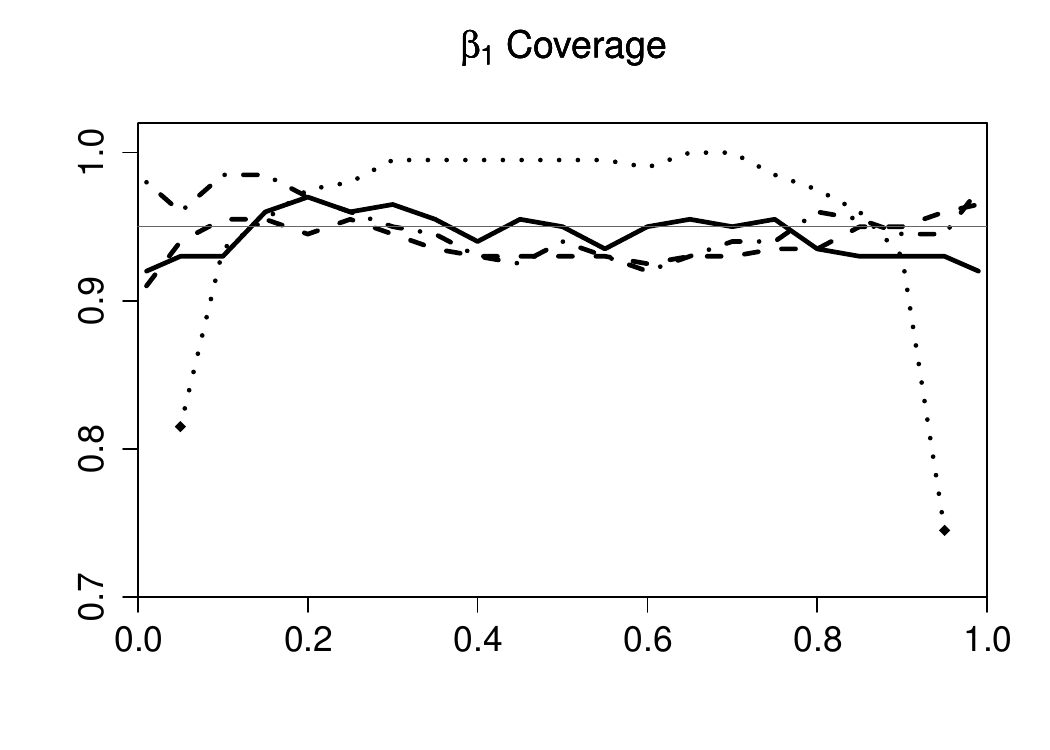} \\[-1.5cm]
    \includegraphics[width=8cm,height=7.5cm,angle=-90]{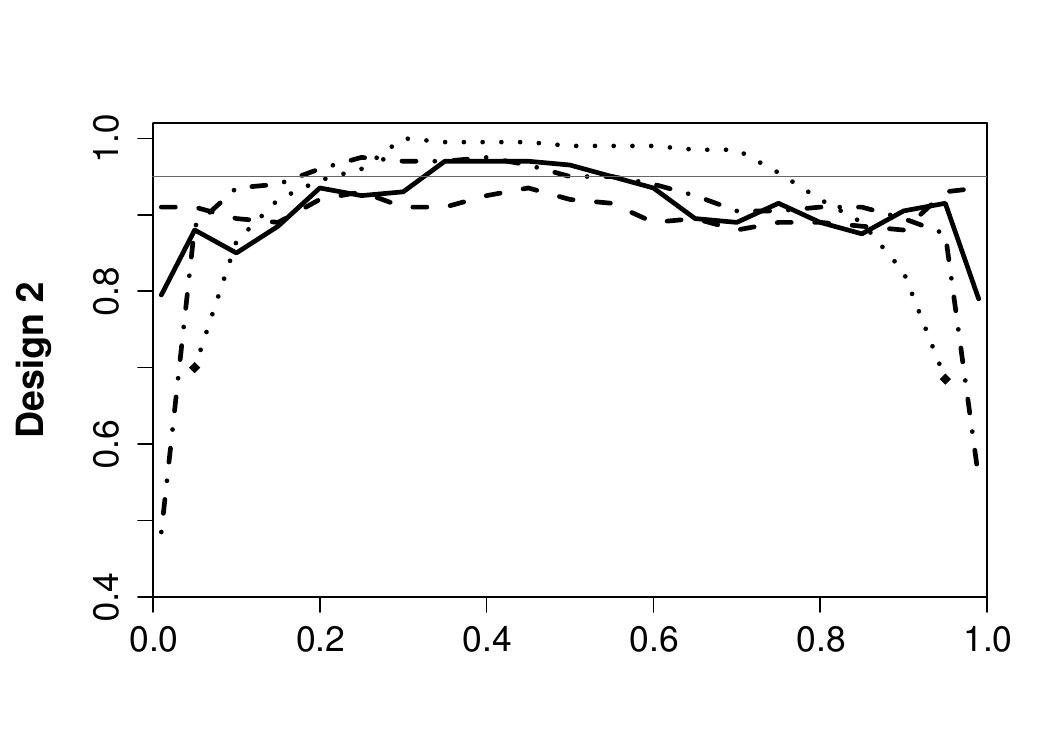}
    \includegraphics[width=8cm,height=7.5cm,angle=-90]{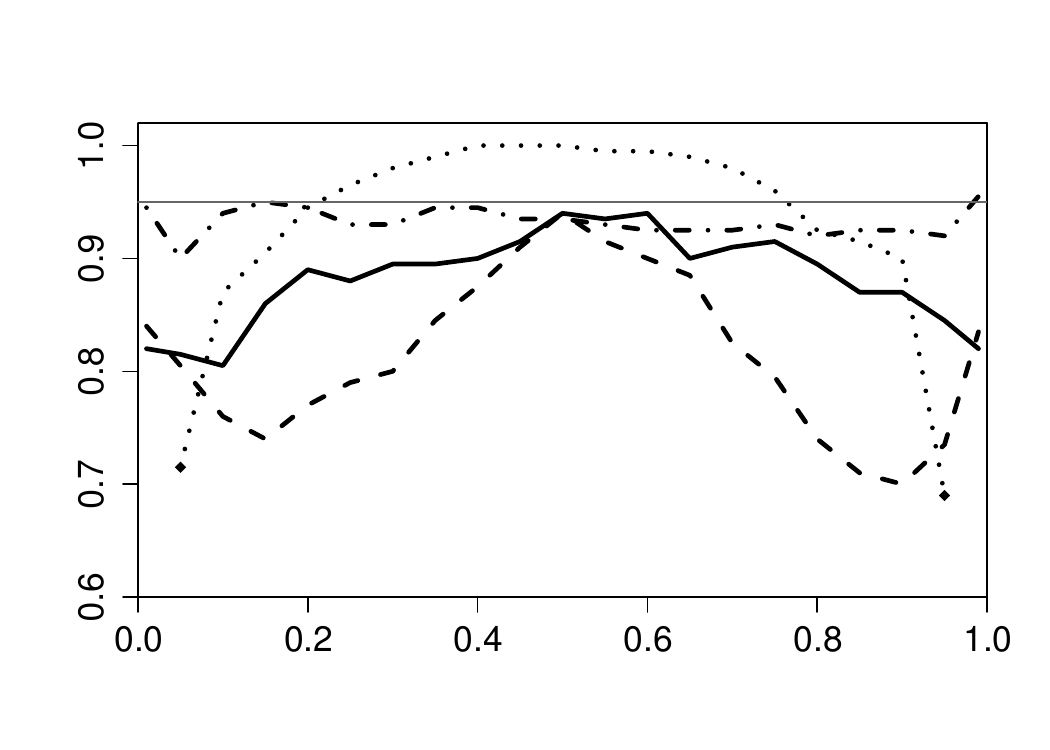} \\[-1.5cm]
    \includegraphics[width=8cm,height=7.5cm,angle=-90]{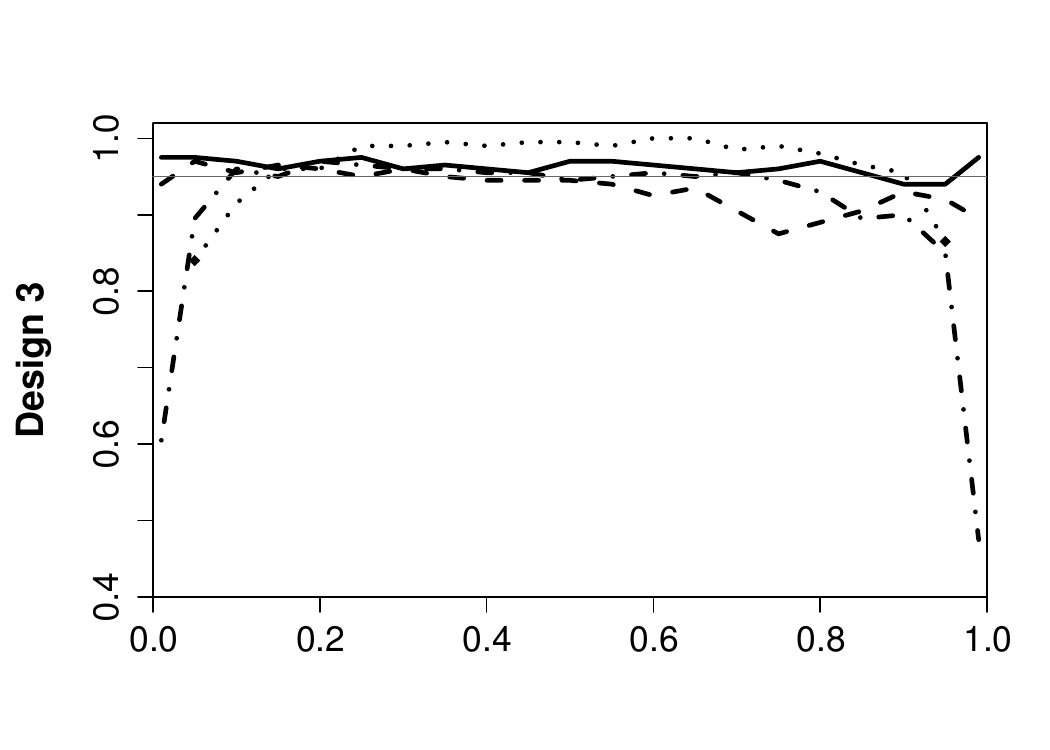}
    \includegraphics[width=8cm,height=7.5cm,angle=-90]{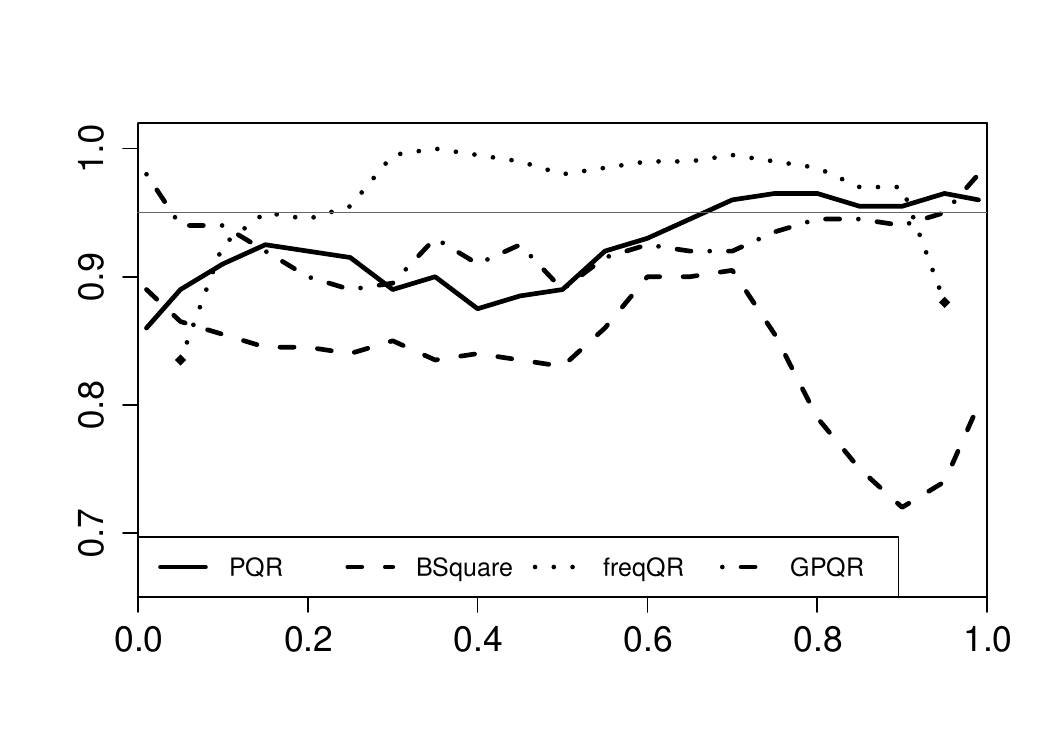}
    \caption{$95\%$ Coverage probabilities for parameters $\beta_0$ (left) and $\beta_1$ (right) at $\tau=0.01,0.05,0.1,\dots,0.95, 0.99$. freqQR coverage probabilities at $\tau=0.01 \text{ and } 0.99$ are not available, so this curve is truncated at $\tau=0.05 \text{ and } 0.95$ (diamond points).}
    \label{fig:sim_linear_cover}
\end{figure}

Figure \ref{fig:sim_linear_cover} shows $95\%$ coverage probabilities at $\tau=0.01,0.05,0.1,\dots,0.95, 0.99$ for the univariate designs. However, freqQR confidence intervals for the parameters at $\tau=0.01 \text{ and } 0.99$ are not available for this sample size, so freqQR results in Figure \ref{fig:sim_linear_cover} are truncated at $\tau=0.05 \text{ and } 0.95$, and highlighted by diamond endpoints.

GPQR has poor coverage probabilities for $\beta_0$ for extreme quantiles, for which the method also presented high RMSE (Figures \ref{fig:sim_linear_mse} and \ref{fig:sim_linear_cover}). Prior complexity naturally compromises model interpretability and usage, which is a disadvantage of this approach. Prior information might be affecting estimation here, although default settings were used. From Figure \ref{fig:sim_linear_cover}, we can also see that freqQR coverages are generally too wide for middle quantiles and too narrow at the extremes (for $\tau = 0.05 \text{ and } 0.95$, as the more extremes are not available). The BSquare approach performed poorly for some of the parameters. PQR has, in general, nice coverage probabilities compared to the alternative approaches. 

For the multivariate design $4$, we considered the estimation at quantile levels $\tau =0.01, 0.05, 0.50$ with $N=350$ samples. PQR was fitted based on $150.000$ MCMC draws and burn-in of $50.000$. For the other methods, previous configurations were adopted. RMSE and coverage results are presented in Tables \ref{tab:mse_des4} and \ref{tab:cover_des4}, respectively.

% Table created by stargazer v.5.2 by Marek Hlavac, Harvard University. E-mail: hlavac at fas.harvard.edu
% Date and time: Thu, Apr 28, 2016 - 12:45:37
% Requires LaTeX packages: dcolumn 
\begin{table}[!htbp] \centering 
  \caption{RMSE ($\times 100$) for design 4} 
  \label{tab:mse_des4} 
\begin{tabular}{@{\extracolsep{5pt}} D{.}{.}{-2} D{.}{.}{-2} D{.}{.}{-2} D{.}{.}{-2} D{.}{.}{-2} D{.}{.}{-2} D{.}{.}{-2} } 
\\[-1.8ex]\hline 
\hline \\[-1.8ex] 
\multicolumn{1}{c}{} & \multicolumn{1}{c}{$\beta_0$} & \multicolumn{1}{c}{$\beta_1$} & \multicolumn{1}{c}{$\beta_2$} & \multicolumn{1}{c}{$\beta_3$} & \multicolumn{1}{c}{$\beta_4$} & \multicolumn{1}{c}{$\beta_5$} \\ 
\hline \\[-1.4ex]
\multicolumn{1}{l}{$\tau=0.50$} &  &  &  &  &  &   \\[0.2ex]
\multicolumn{1}{l}{\hspace{0.3cm} PQR} & 13.00 & 24.74 & 21.15 & 20.74 & 20.29 & 19.19 \\ 
\multicolumn{1}{l}{\hspace{0.3cm} BSquare} & 16.13 & 28.20 & 26.52 & 19.85 & 18.91 & 19.00 \\ 
\multicolumn{1}{l}{\hspace{0.3cm} freqQR} & 13.64 & 22.37 & 23.47 & 21.71 & 21.14 & 22.05 \\ 
\multicolumn{1}{l}{\hspace{0.3cm} GPQR} & 13.37 & 24.86 & 23.00 & 22.61 & 21.84 & 21.03 \\   [1ex]
\multicolumn{1}{l}{$\tau=0.05$} &  &  &  &  &  &   \\[0.2ex]
\multicolumn{1}{c}{\hspace{0.3cm} PQR} & 21.31 & 30.61 & 35.89 & 32.16 & 34.63 & 31.55 \\ 
\multicolumn{1}{c}{\hspace{0.3cm} BSquare} & 22.62 & 65.06 & 83.90 & 19.78 & 19.51 & 19.32 \\ 
\multicolumn{1}{c}{\hspace{0.3cm} freqQR} & 21.86 & 37.57 & 39.69 & 35.90 & 38.68 & 37.01 \\ 
\multicolumn{1}{c}{\hspace{0.3cm} GPQR} & 20.85 & 32.44 & 36.93 & 32.62 & 34.58 & 30.18 \\ [1ex]
\multicolumn{1}{l}{$\tau=0.01$} &  &  &  &  &  &   \\ [0.2ex]
\multicolumn{1}{c}{\hspace{0.3cm} PQR} & 32.73 & 40.03 & 48.45 & 39.74 & 43.11 & 38.49 \\ 
\multicolumn{1}{c}{\hspace{0.3cm} BSquare} & 65.67 & 72.44 & 91.51 & 19.77 & 19.50 & 19.31 \\
\multicolumn{1}{c}{\hspace{0.3cm} freqQR} & 39.31 & 52.41 & 57.09 & 52.83 & 54.29 & 51.39 \\ 
\multicolumn{1}{c}{\hspace{0.3cm} GPQR} & 45.67 & 50.27 & 57.16 & 48.12 & 50.34 & 46.58 \\ 
\hline \\[-1.8ex] 
\end{tabular} 
\end{table}

% Table created by stargazer v.5.2 by Marek Hlavac, Harvard University. E-mail: hlavac at fas.harvard.edu
% Date and time: Thu, Apr 28, 2016 - 13:47:47
% Requires LaTeX packages: dcolumn 
\begin{table}[!htbp] \centering 
  \caption{$95\%$ Coverage probabilities for design 4} 
  \label{tab:cover_des4} 
\begin{tabular}{@{\extracolsep{5pt}} D{.}{.}{-2} D{.}{.}{-2} D{.}{.}{-2} D{.}{.}{-2} D{.}{.}{-2} D{.}{.}{-2} D{.}{.}{-2} } 
\\[-1.8ex]\hline 
\hline \\[-1.8ex] 
\multicolumn{1}{c}{} & \multicolumn{1}{c}{$\beta_0$} & \multicolumn{1}{c}{$\beta_1$} & \multicolumn{1}{c}{$\beta_2$} & \multicolumn{1}{c}{$\beta_3$} & \multicolumn{1}{c}{$\beta_4$} & \multicolumn{1}{c}{$\beta_5$} \\ 
\hline \\[-1.4ex]
\multicolumn{1}{l}{$\tau=0.50$} &  &  &  &  &  &   \\[0.2ex]
\multicolumn{1}{c}{\hspace{0.3cm} PQR} & 0.94 & 0.90 & 0.95 & 0.95 & 0.96 & 0.96 \\ 
\multicolumn{1}{c}{\hspace{0.3cm} BSquare} & 0.88 & 0.82 & 0.90 & 0.96 & 0.94 & 0.94 \\ 
\multicolumn{1}{c}{\hspace{0.3cm} freqQR} & 0.98 & 1.00 & 0.99 & 0.98 & 1.00 & 0.98 \\ 
\multicolumn{1}{c}{\hspace{0.3cm} GPQR} & 0.96 & 0.86 & 0.89 & 0.92 & 0.95 & 0.92 \\ [1ex]
\multicolumn{1}{l}{$\tau=0.05$} &  &  &  &  &  &   \\[0.2ex] 
\multicolumn{1}{c}{\hspace{0.3cm} PQR} & 0.96 & 0.98 & 0.94 & 0.98 & 0.96 & 0.96 \\ 
\multicolumn{1}{c}{\hspace{0.3cm} BSquare} & 0.96 & 0.19 & 0.09 & 0.96 & 0.94 & 0.94 \\ 
\multicolumn{1}{c}{\hspace{0.3cm} freqQR} & 0.90 & 0.88 & 0.86 & 0.89 & 0.85 & 0.90 \\ 
\multicolumn{1}{c}{\hspace{0.3cm} GPQR} & 0.92 & 0.92 & 0.92 & 0.96 & 0.92 & 0.94 \\ [1ex]
\multicolumn{1}{l}{$\tau=0.01$} &  &  &  &  &  &   \\ [0.2ex]
\multicolumn{1}{c}{\hspace{0.3cm} PQR} & 0.97 & 0.96 & 0.91 & 0.96 & 0.96 & 0.98 \\ 
\multicolumn{1}{c}{\hspace{0.3cm} BSquare} & 0.72 & 0.14 & 0.08 & 0.96 & 0.94 & 0.94 \\ 
\multicolumn{1}{c}{\hspace{0.3cm} freqQR} & 0.41 & 0.52 & 0.46 & 0.47 & 0.48 & 0.50 \\ 
\multicolumn{1}{c}{\hspace{0.3cm} GPQR} & 0.76 & 0.96 & 0.97 & 0.97 & 0.98 & 0.97 \\ 
\hline \\[-1.8ex] 
\end{tabular} 
\end{table}

BSquare had issues in estimating the parameters for this multivariate design. In particular $\beta_0, \beta_1 \text{ and } \beta_2$ presented high RMSE's and low coverages, as shown in Tables \ref{tab:mse_des4} and \ref{tab:cover_des4}. 
In fact, the estimated quantile planes corresponding to different quantile levels were generally parallel, which obviously impacted the estimation of all parameters that vary with $\tau$. This drawback of the non-crossing constraints imposed in \shortciteN{reichs13} often happens for multivariate examples, unless large samples are available so that crossing occurs infrequently.

From Table \ref{tab:mse_des4}, PQR has generally the smallest RMSE, significantly outperforming GPQR at $\tau=0.01$ and also notably better than freqQR for $\tau=0.05\text{ and } \tau=0.01$. Moreover, among all methods, PQR has coverages closest to the nominal level. As noted before, freqQR has coverages consistently above the nominal level at $\tau=0.50$ and below it at the extremes ($\tau=0.05 \text{ and } \tau=0.01$). Again GPQR has poor coverage for $\beta_0$ for extreme quantiles. 

Note that PQR has great performance despite the small number of pyramid levels ($M=2$, $\tau =0.01, 0.05, 0.50$). Indeed, increasing $M$ does not significantly affect the results, corroborating the proximity between the least false and true parameter values. 

%\textcolor{red}{Naturally, simultaneous quantile regression is computationally more demanding than individual quantile fitting, and it is specially advisible when sample sizes are small and crossing occurs. When sample sizes are large, the simple minimization problem proposed by \citeN{KoenkerBasset1978} has great coverages and generally small errors, as shown in \citeN{yang2015}. Therefore, we concentrate our attention here in the small sample size problem. Nevertheless, one can also use this framework to fit larger datasets. }

We have restricted our simulations studies to relatively small sample sizes since in large samples, the simple minimization problem proposed by \citeN{KoenkerBasset1978} has great coverages and generally small errors, as shown in \citeN{yang2015}.  Under our framework, 
we would expect MCMC to converge faster since in large samples crossing of quantiles is less likely to occur and hence slow down
the MCMC sampling algorithm. In terms of computational cost, around 80\% of the computational overhead is attributable to the likelihood calculation.
%MCMC runtimes here are almost entirely (generally above 80\%) devoted to the computation of the likelihood, 
This is mostly due to the indicator function in Equation \eqref{eqnLik}.  For example, the times it takes to compute one likelihood using non-optimised codes, for multivariate design 4, are $3\times10^{-4}$ seconds for $N=350$ and $2\times10^{-3}$ seconds for $N=3500$, in a 3.6GHz quad-core Intel i7-4790k CPU.  However, the likelihood evaluations are highly parallelisable and runtimes can be reduced dramatically.
The additional computational burden with increasing number of covariates is insignificant compared to the cost of likelihood evaluations,
but of course it causes a linear increase in the number of parameters updated at each iteration, $(P+1)*(T+2)$.

\section{Real examples} \label{sec:realex}
In this section, we illustrate the proposed method on two publicly available real datasets, one involving extremal quantile modelling and a censored data analysis involving  a large number of covariates.

\subsection{Extreme quantile modelling}
In extreme value analysis, it is common practice to use the so-called extreme value distributions to make inference on the tails of the distribution of the data. 
Using a parametric model places strong assumptions on the data, but is an attractive approach since data is often scarce in the extremal regions.
However, a long standing issue is the fidelity of the data to the parametric assumptions, see \shortciteN{coles01}. 
We propose in this application to model linear quantiles of extreme data using PQR, that allows us to drop these parametric assumptions, but instead 
use the information from extreme distributions as prior when centring the quantile process.

Here we will apply PQR to model extreme tropical cyclones. The dataset consists of $82$ observations of cyclones whose wind speed is greater than $96$ knots (kt) threshold, recorded in the US coast from 1899 to 2006 (this is an updated version of the data analysed in \shortciteNP{jagger09} which included $79$ cyclones; the update is available in the authors' webpage).  \shortciteN{jagger09} considers that these data follow a Generalized Pareto Distribution (GPD),  with cdf given by
$$ G(y) = 1 - [1 + \xi (y-\mu)/ \sigma ]_{+}^{- 1/ \xi} \, ,$$
where $ (h)_+ = max(h,0)$, $\mu$ is the fixed threshold, and $\sigma>0$ and $\xi$ are the scale and shape parameters respectively.

\begin{figure}[!ht]
    \centering
    \subfloat[PQR fitted quantile planes]{\includegraphics[height=6cm,width=6cm,angle=-90]{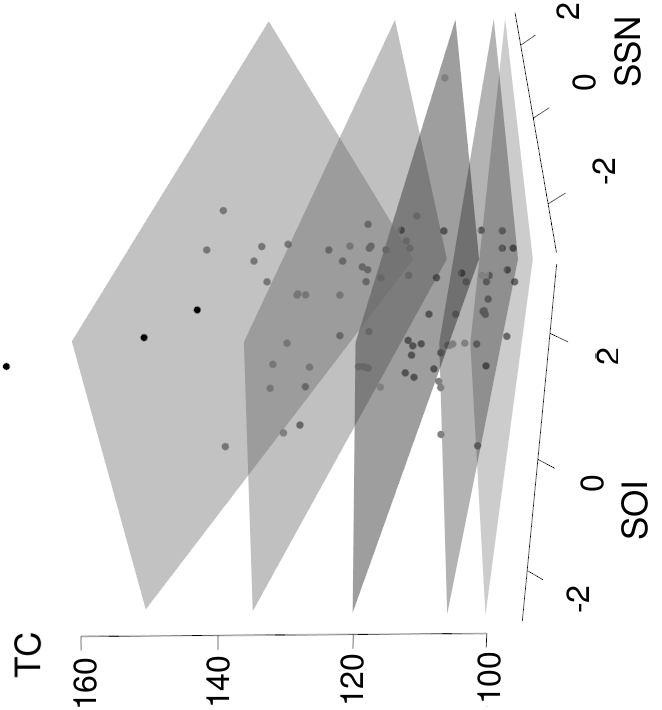}} \hspace{0.5cm}
    \subfloat[Intercept]{\includegraphics[height=6cm,width=6cm,angle=-90]{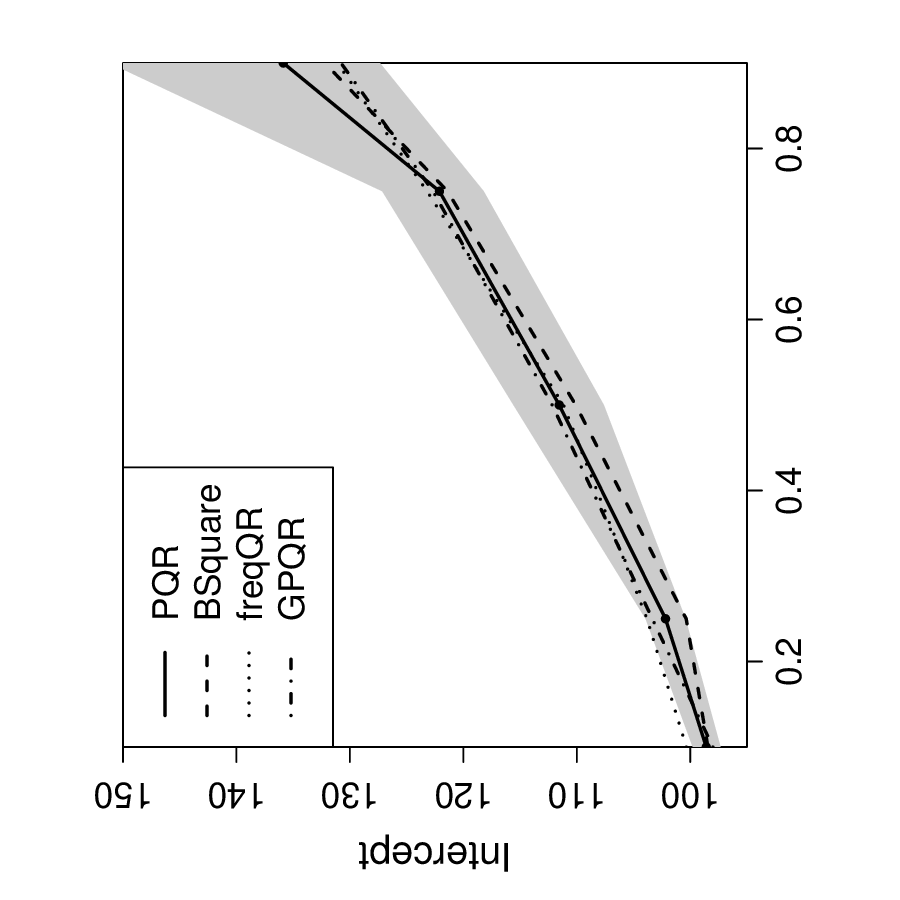}} \\
    \vspace{-0.6cm}
    \subfloat[SOI covariate]{\includegraphics[height=6cm,width=6cm,angle=-90]{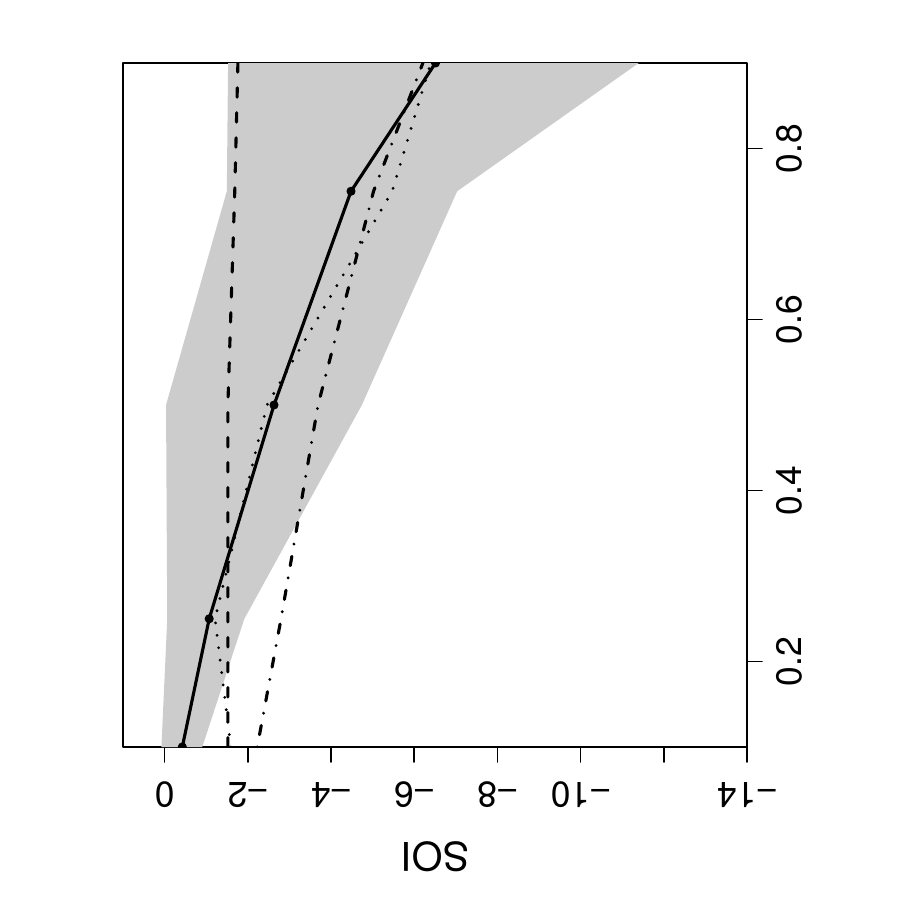}}   \hspace{0.5cm}
    \subfloat[SSN covariate]{\includegraphics[height=6cm,width=6cm,angle=-90]{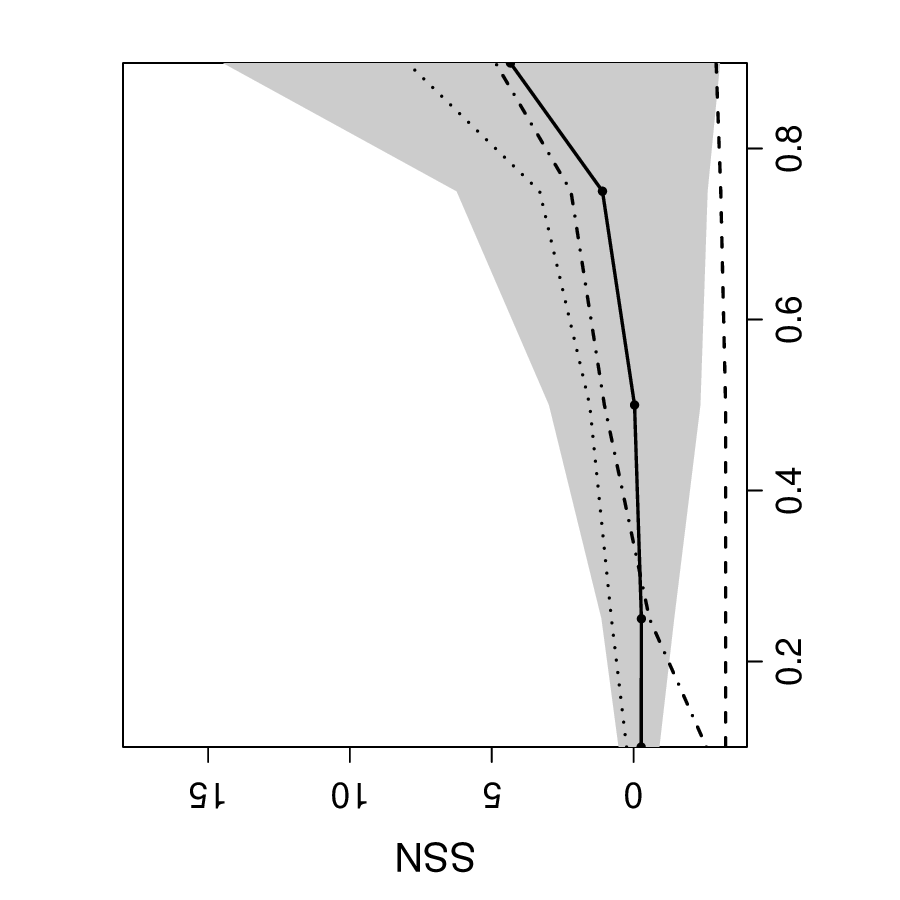}}
    \caption{Estimation of extreme tropical cyclone wind speed (TC, in knots) at $\tau=0.10,0.25,0.50,0.75,0.90$. (a) PQR fitted quantile planes. (b-d) Parameter estimates using PQR (solid line), BSquare (dashed line), freqQR (dotted line) and GPQR (dash-dotted line). The grey shading indicates $90\%$ confidence interval for PQR.}
    \label{fig:cyc}
\end{figure}

Therefore, we consider fitting PQR using GPD as the quantile process centring distribution. Similarly to the Gaussian case  we assume here that the unknown parameters ($\sigma$, $\xi$) change linearly in $x$. Furthermore, we use $\text{Gamma}(0.001,0.001)$ and $U(0,1000)$ as hyperpriors for $\sigma$ and $\xi$, respectively. Following \shortciteN{jagger09}, we model extreme tropical cyclone (TC) wind speed quantiles at $\tau=0.10,0.25,0.50,0.75,0.90$ as a function of the Southern Oscillation Index (SOI) and the sunspot number (SSN), both averaged over August-October and standardised. We used for the estimation $60.000$ MCMC draws and burn-in of $10.000$. Figure \ref{fig:cyc}(b-d) presents the parameter estimates and $90\%$ confidence interval for PQR, obtained as the upper and lower 0.05 sample quantiles of the posterior samples. For comparison, BSquare, freqQR and GPQR estimates are also indicated.

As illustrated in Figure \ref{fig:cyc}, wind speed increases with decreasing SOI, which is expected as small SOI is associated with El Nino warming events, which in turn favour extreme cyclones, as explained in \shortciteN{jagger09}. As in \shortciteN{jagger09}, SSN is generally positive associated with extreme winds, but this is not a statistically significant association.  From SOI parameter estimates' plot, we can also see that PQR provides smoother and nicer estimates than freqQR, which lack borrowing strenght from the neighbours $\tau$. Due to the rigid non-crossing constraints, BSquare parameter estimates SOI and SSN are constant and lie mostly outside PQR $90\%$ confidence interval. GPQR produced generally smaller estimates than PQR and freqQR in the SOI parameter across the quantile levels.  For the SSN parameter, GPQR produced smaller estimates only in the lower quantiles, while the other estimates largely agree with PQR and freqQR. Therefore, in the event that the data truly follow the GPD distribution, by placing priors centered on this distribution, we retain some advantages of using the parametric model for inference, and should perform better than models that cannot incorporate this information. However, in the case where data deviates from GPD, the pyramid quantile framework can correct for this misspecification with increasing data. So the ability of the pyramid quantiles to place informative priors allows us to fully take advantage of the Bayesian inferential framework.

\subsection{Analysis of censored data}
Regression with large numbers of covariates poses additional computational challenges for the proposed method. 
Here we consider the University of Massachusetts Aids Research Unit IMPACT study data (UIS) available in the quantreg package in R, from \shortciteN{hosmerl98}, and analysed by \shortciteN{portnoy03}, \shortciteN{reichs13} and \shortciteN{yang2015} using quantile regression. For this analysis with right censoring, the  log-likelihood  is now the sum over $i=1,\dots,n$ of
$$
(1-c_i)\log f(y_i|{\bf x}_i) + c_i\log (1-F(y_i|{\bf x}_i))
$$
where $f(y_i|{\bf x}_i)$ is given by Equation \ref{eqnLik}, where $F(y_i|{\bf x}_i)$ is the corresponding CDF and where $c_i$ is the censoring status (1=right censored, 0=otherwise).

The dataset contains records for 575 observations, we estimated the conditional quantiles for the logarithm of the time to return to drug use ($Y$) as linear functions of 8 predictors, 
BECK (a depression score), FRAC (a compliance factor), AGE (age at enrollment), TREAT (current treatment assignment, 1= Long course, 0=Short course), NDT (number of previous drug treatment), RACE (1=Non-white, 0=White), IV3 (recent intravenous drug use, 1=Yes, 0=No), SITE (treatment site). All variables were scaled by subtracting their mean and dividing by their range.
We fit the quantile levels $\tau=0.1, 0.2,\ldots, 0.9$ using 9 quantile pyramids and the Gaussian centring distribution, this amounts to a problem with 99 parameters. For high dimensions,  the strategy described in Section \ref{mcmc} requires several modifications. 

Firstly, existing off-the-shelf convex hull algorithms encounter memory problems for dimensions higher than 7 or 8. 
Here our strategy is to compute the convex hull of the data expressed in the space given by their leading 5 or 6 principal components, 
then to choose the remaining vertices by random sampling. We trial 500 random samples in this fashion, and select the pyramid locations that has the
maximum distance between the quantile levels over the $P+1$  locations. Non-crossing constraint is then verified at all data points.

In higher dimensions, we also have noticed that well placed pyramid locations can greatly improve the MCMC mixing  since the parameters are often highly correlated. 
%, often these parameters are also highly correlated, which can lead to slow MCMC mixing. 
For the current problem, we perform several parallel runs, each corresponding to a different set of pyramid locations, and choose the best mixing chain. 
More precisely for each chain we performed a trial MCMC run of 20.000 of standard MCMC, updating one parameter at a time, with tuning of proposal variance to obtain acceptance probability of roughly 0.44 for each parameter. This step allows us to learn the covariance structure of the parameters. 

The next stage of MCMC incorporates the information learned in the first stage, by blocking variables into separate groups at each quantile level (over covariates) and  groups at each covariate level (over quantiles), as well as  blocking all the centring parameters $\mu^p$ in one block and all the variance parameters $\sigma^p$ in another. At each iteration of the MCMC, all blocks of the quantiles are updated once, followed by the blocks for $\mu^p$ and $\sigma^p$. The blocks are  updated using the learned covariance matrix from the first stage, and a random walk proposal with Gaussian and truncated Gaussian respectively.  For each quantile block, we iteratively updated each component parameter within the block, by first updating one parameter independently, using the proposal strategy of Section \ref{mcmc}, and then updating the following parameters of the block using their conditional distribution and the covariance structure. Again, non-crossing is verified at each data point as we update each parameter. We found that adding this second MCMC run tend to provide more reliable MCMC output that mixes well for most of the pyramid choices. We ran this second stage for 200.000 iterations with 20.000 samples as burn in. 

\begin{figure}[!htbp]
    %\captionsetup[subfigure]{labelformat=empty}
   % \centering
    \includegraphics[width=4cm,height=5cm,angle=0]{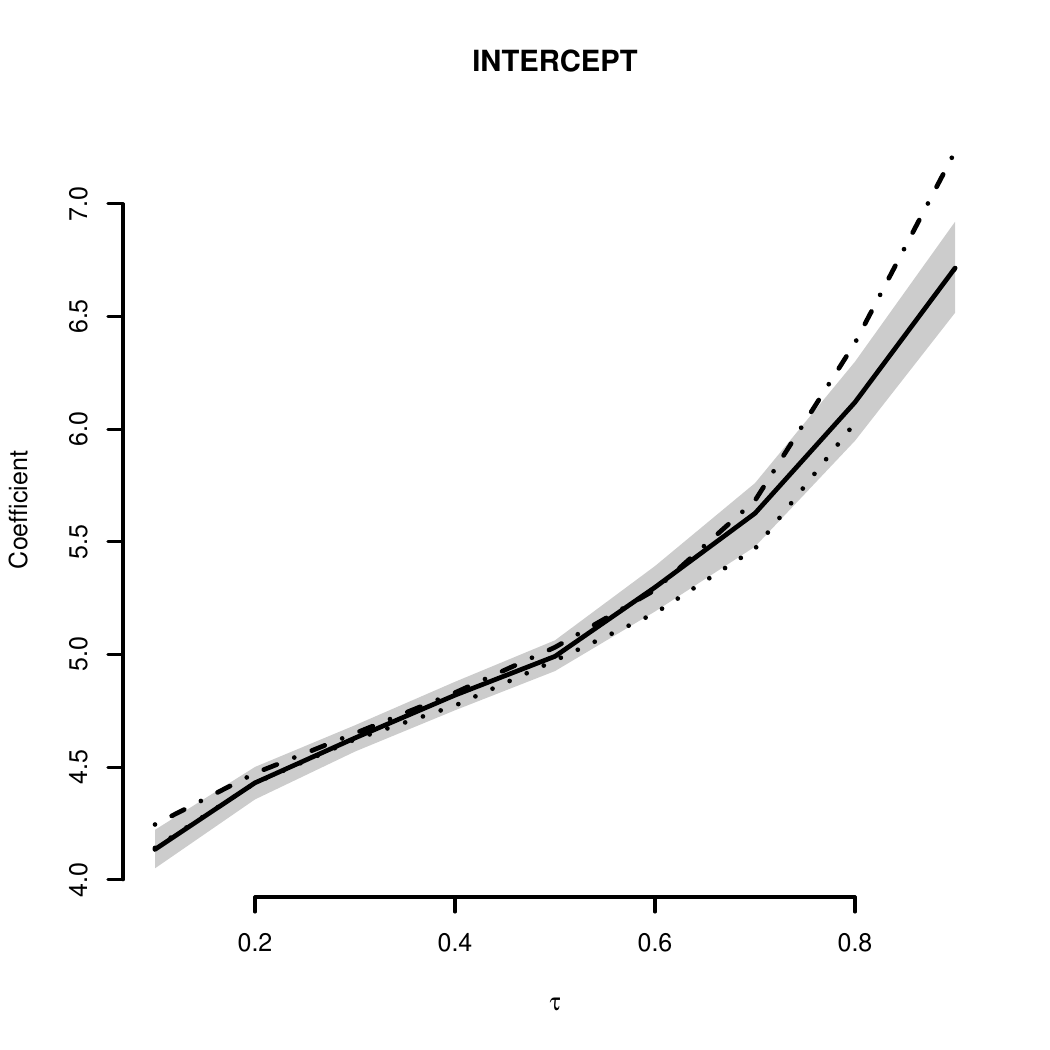}
    \includegraphics[width=4cm,height=5cm,angle=0]{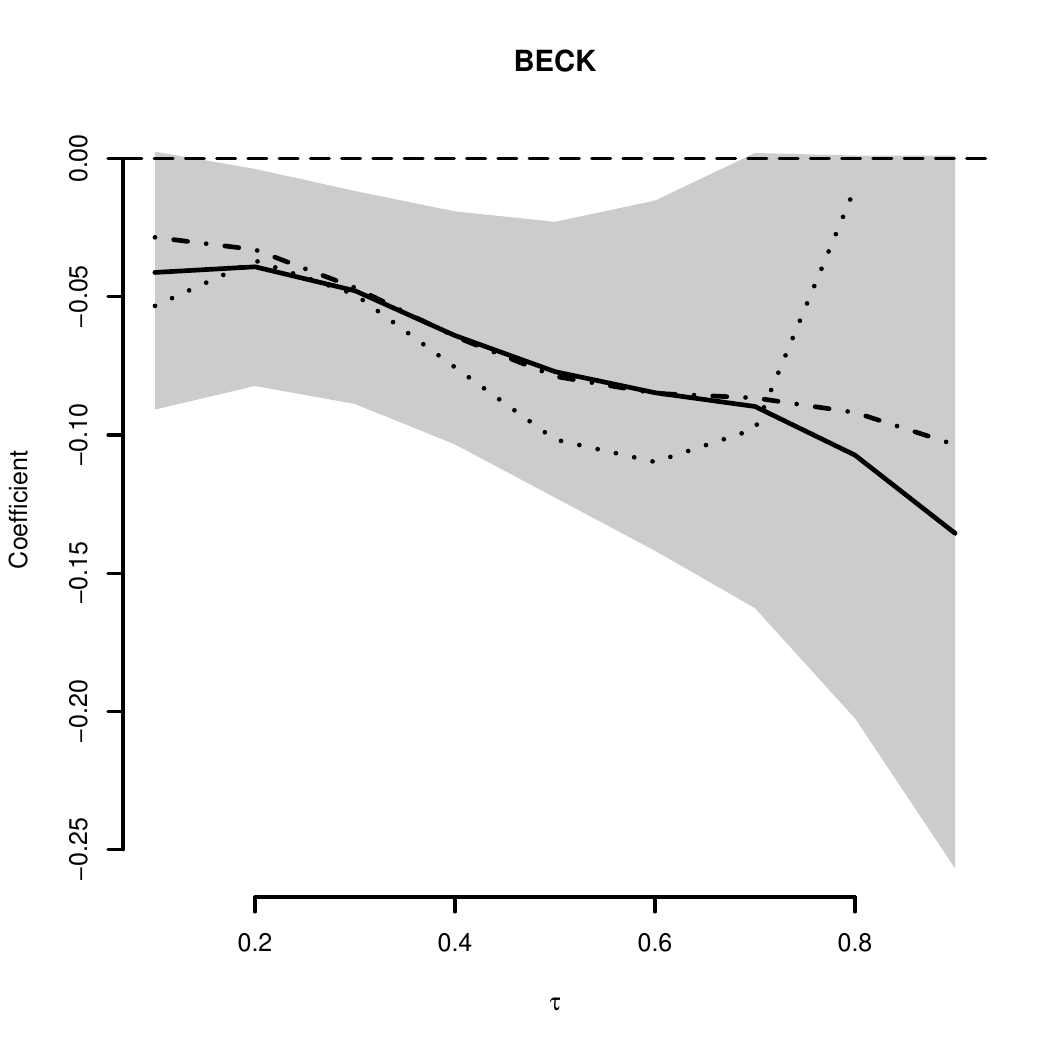} 
        \includegraphics[width=4cm,height=5cm,angle=0]{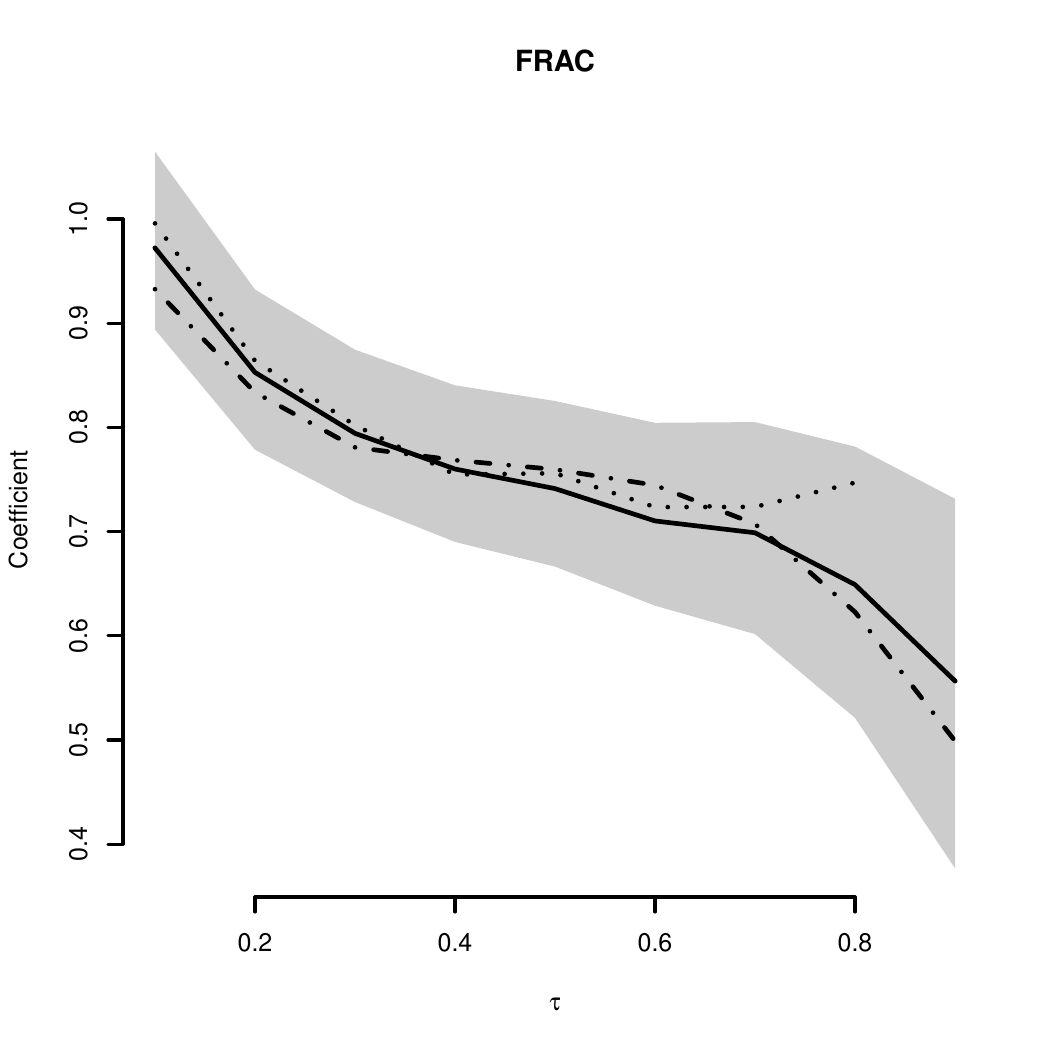}
    \includegraphics[width=4cm,height=5cm,angle=0]{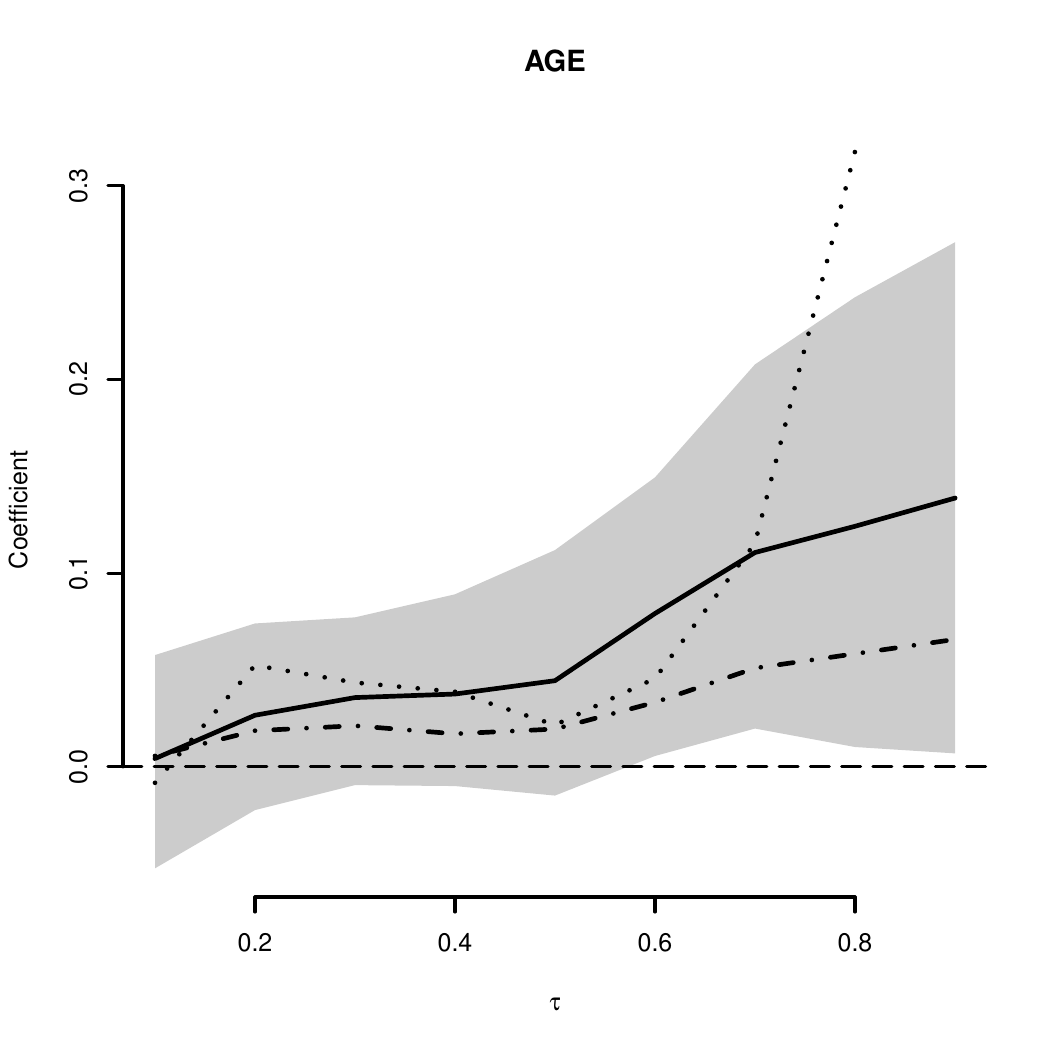} \\[-0cm]
    \includegraphics[width=4cm,height=5cm,angle=0]{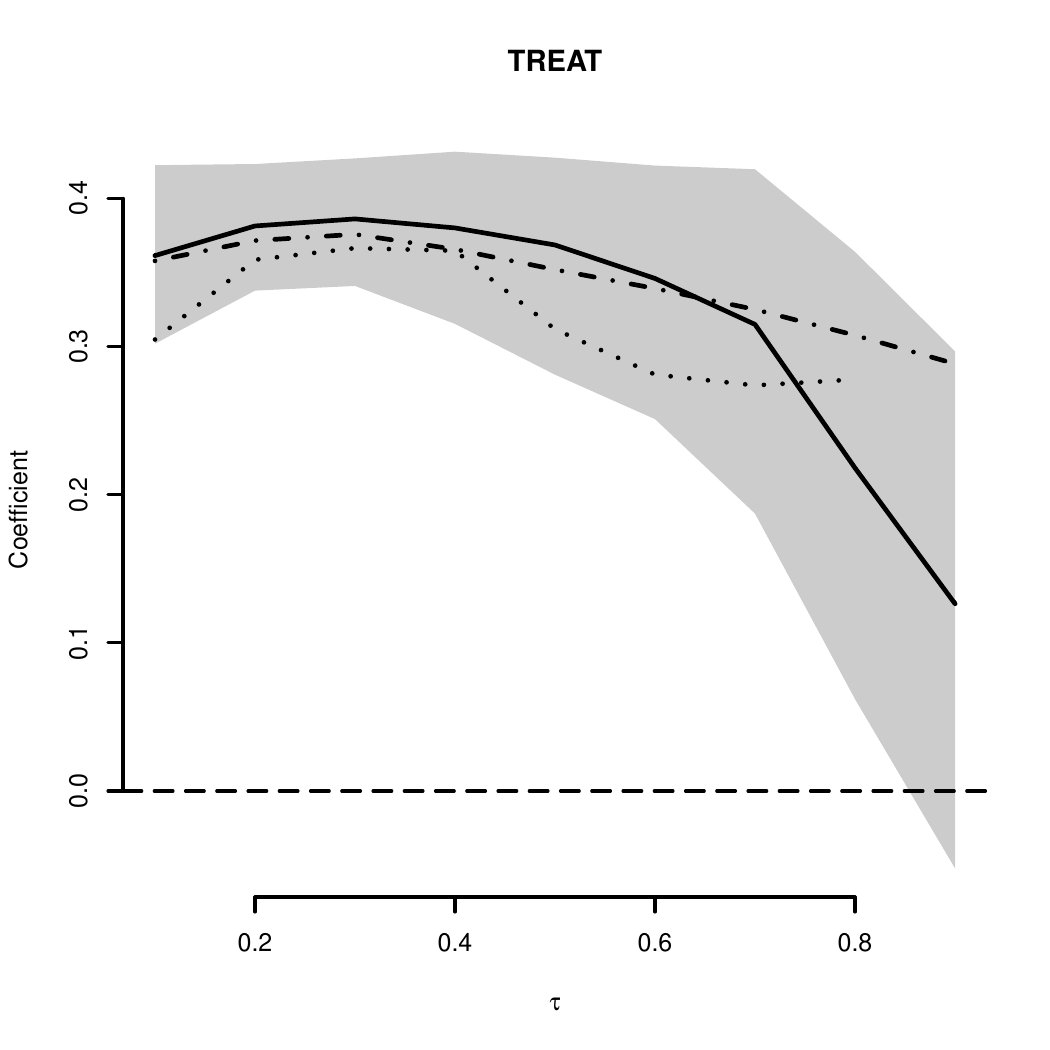}
    \includegraphics[width=4cm,height=5cm,angle=0]{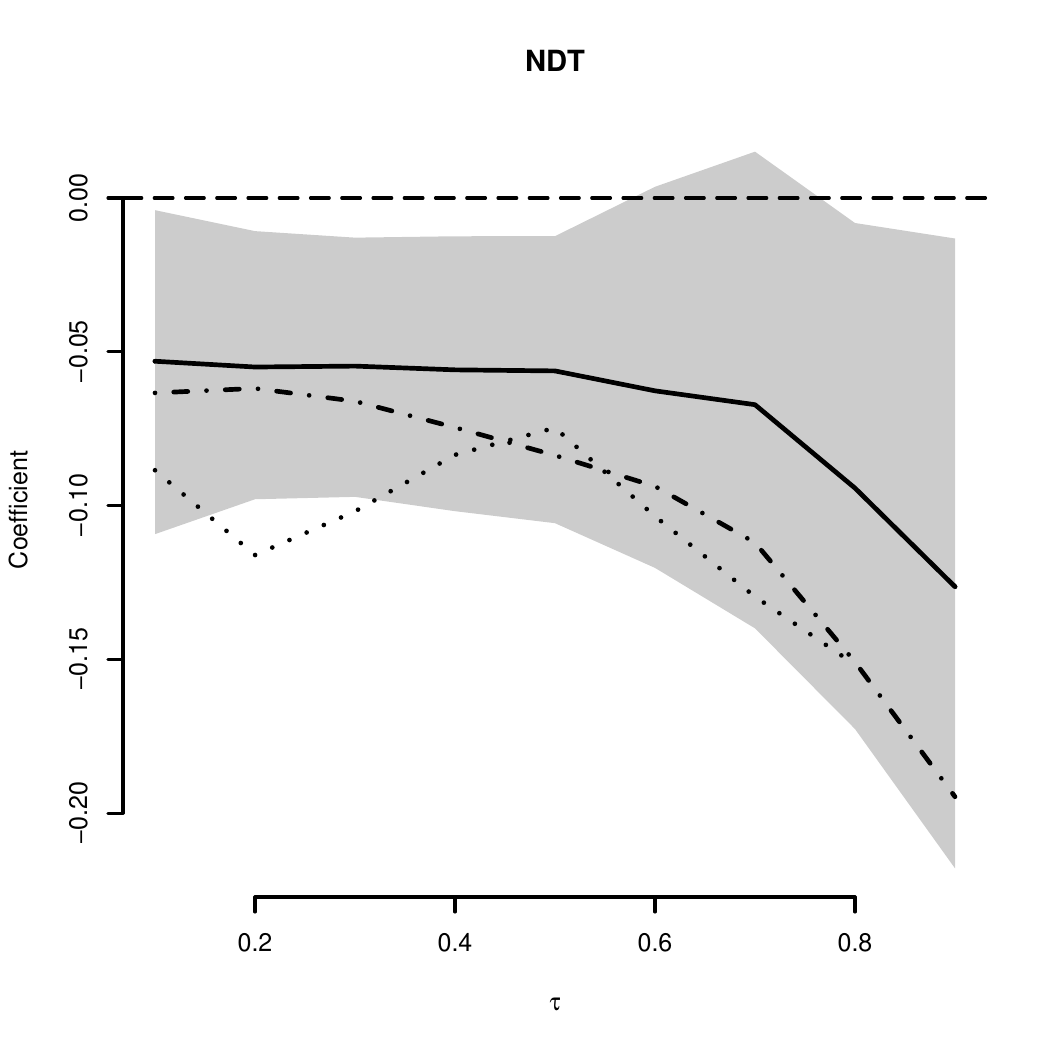}
     \includegraphics[width=4cm,height=5cm,angle=0]{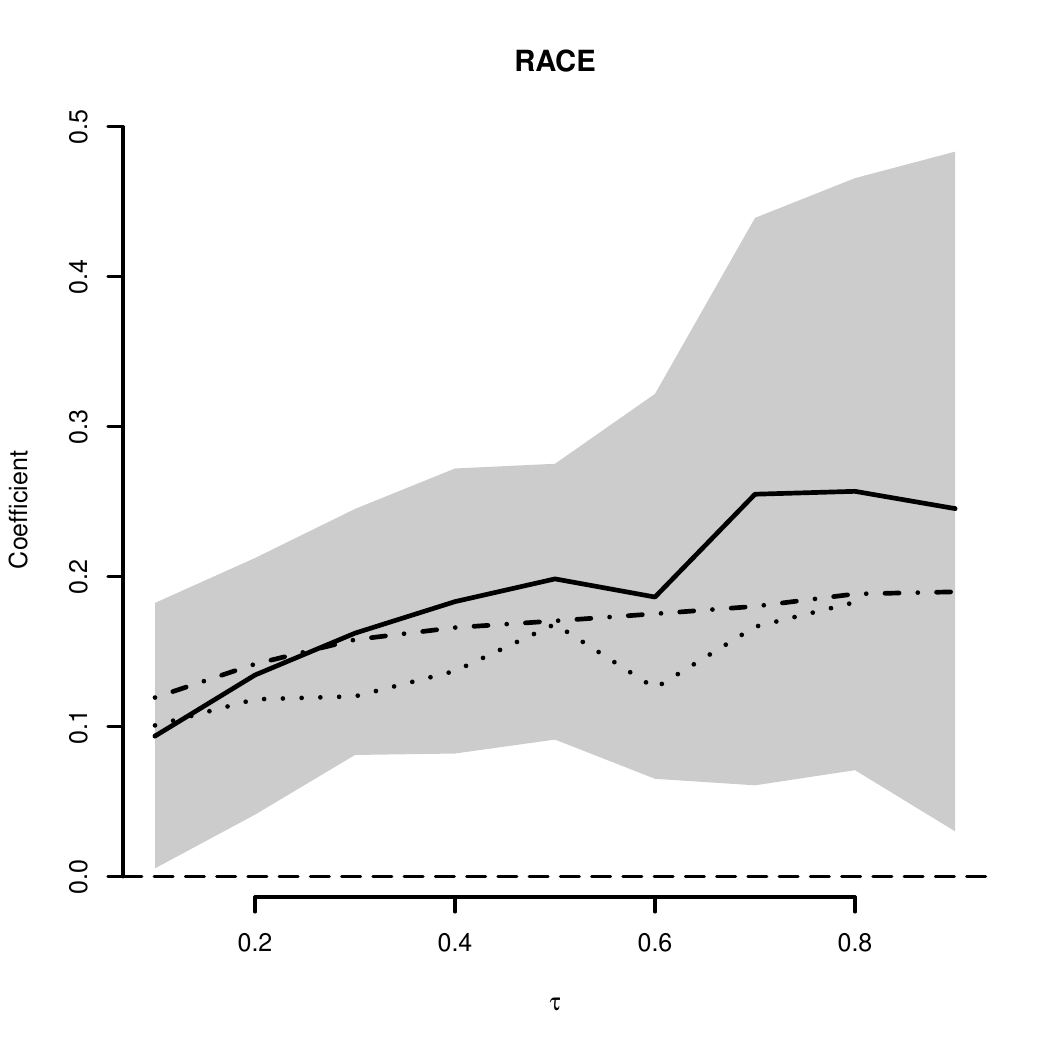}
      \includegraphics[width=4cm,height=5cm,angle=0]{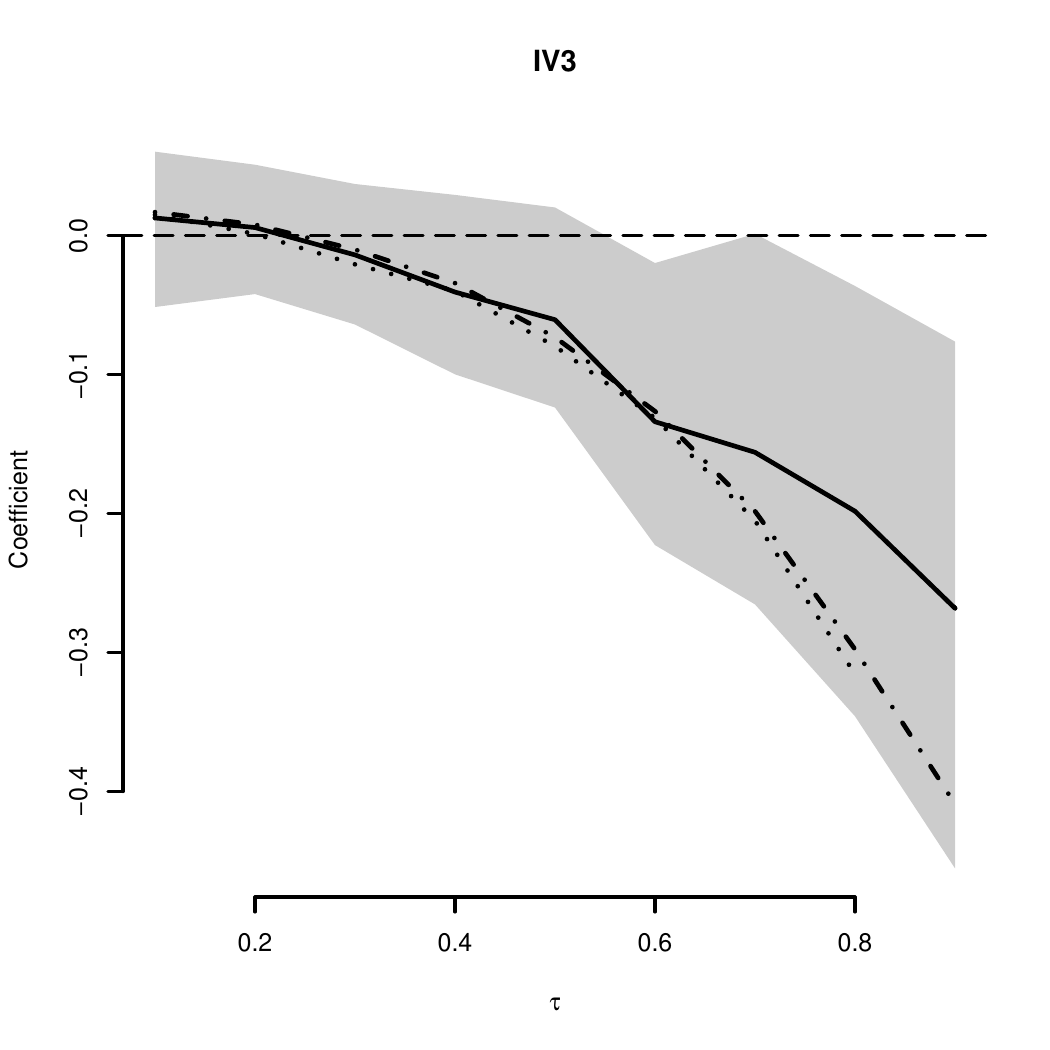}\\[-0cm]
       \includegraphics[width=4cm,height=5cm,angle=0]{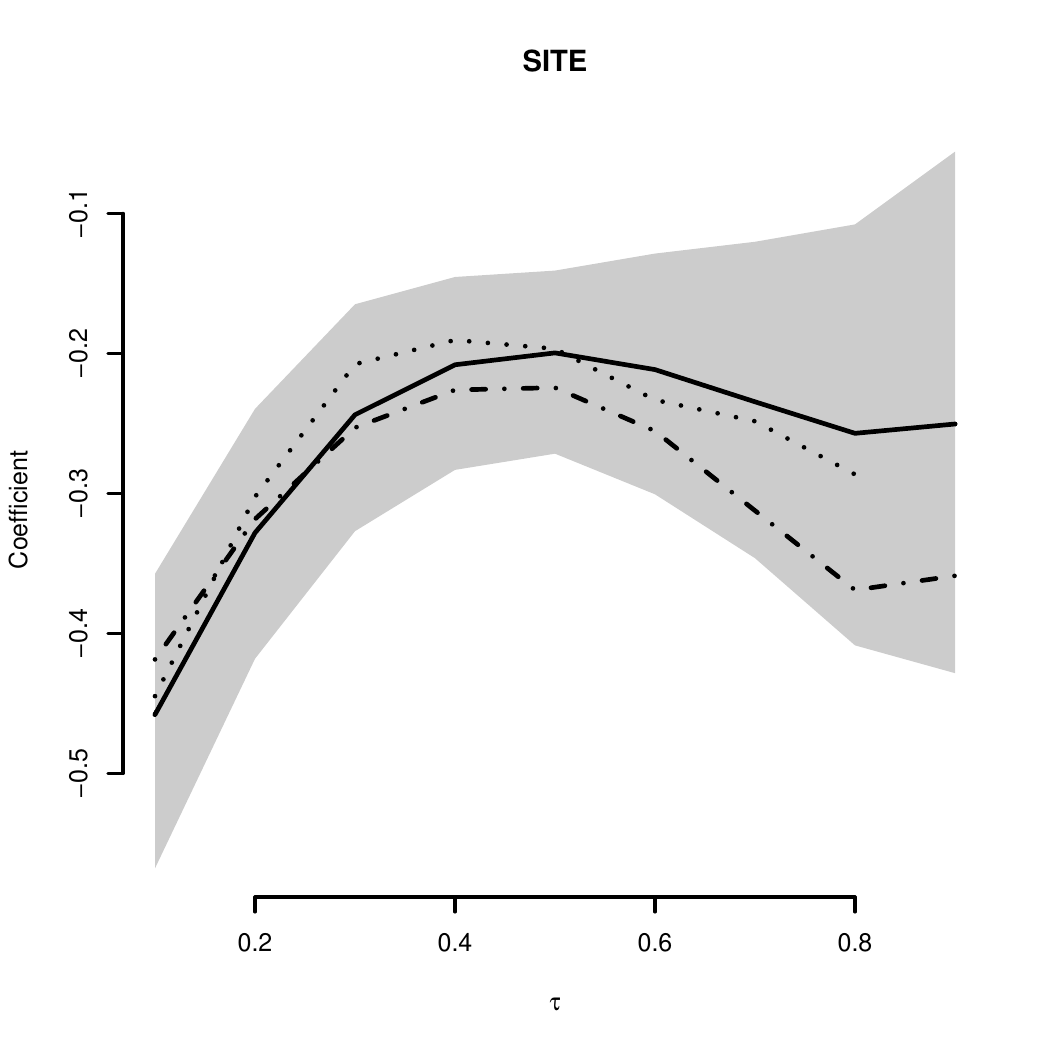}
        \includegraphics[width=4cm,height=5cm,angle=0]{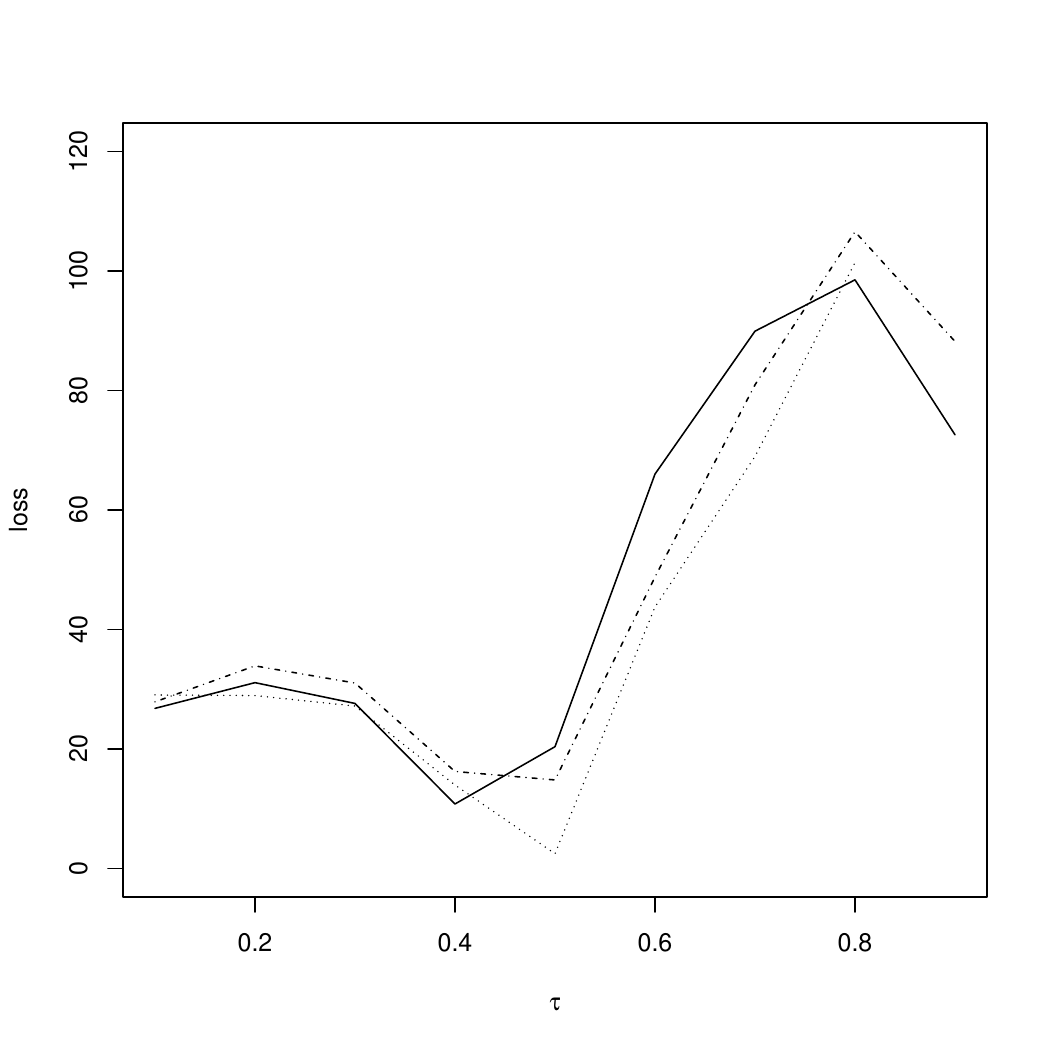}
         \includegraphics[width=4cm,height=5cm,angle=0]{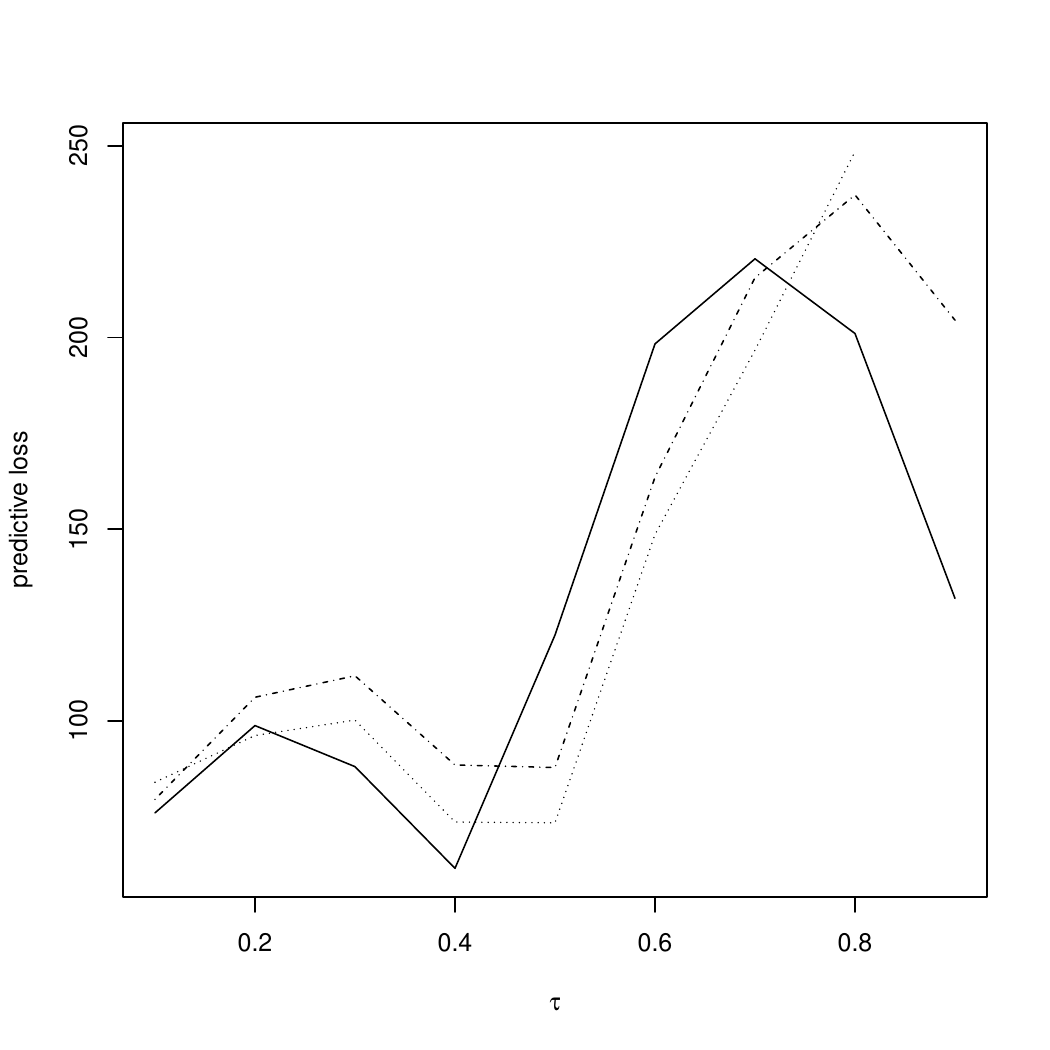}
        \caption{Estimation of regression coefficients for UIS data analysis at $\tau=0.1, 0.2,\ldots, 0.8, 0.9$. Each subplot shows the posterior mean of regression coefficient for the
    respective covariate using PQR (solid line), freqQR (dotted line)  and GPQR (dash-dotted line), dashed line indicates the value at 0. The grey shading indicates $90\%$ confidence interval for PQR. Final plots shows the check loss and predictive check loss.}
    \label{fig:UIS}
\end{figure}

Figure \ref{fig:UIS} show the estimated coefficients over different quantile levels, PQR estimates are given by solid lines. We also implemented the method of \shortciteN{portnoy03} (dotted line) and \shortciteN{yang2015} (dash dotted line). The method of  \shortciteN{portnoy03}  was used to compute the first eight quantile levels, since it does not produce results for quantile level 0.9 or higher. The three methods produced similar results for the lower quantile levels. For a comparison, we computed the check loss (defined in Section 1) at each of the quantile levels $\tau=0.1, 0.2,\ldots, 0.9$, by summing over $\rho_{\tau}(y_i- \widehat{Q}_{Y}(\tau|{\bf X}_i))$, where $y_i$s are the un-censored observations, and ${\bf X}_i$ are the corresponding covariate values. The final subplot in Figure \ref{fig:UIS} shows the computed loss for the three methods. For lower quantiles, there's little difference, whereas the method of \shortciteN{portnoy03} is better for moderate to high quantiles, they do not produce estimates for very high quantiles, nor do they ensure non-crossing. PQR out-performs the other two methods in terms of check loss for higher quantiles, see middle figure in the last row of Figure \ref{fig:UIS}. A similar result is seen in the predictive check loss, when we used 10\% of the data as test data,  see last figure in Figure \ref{fig:UIS}, where the out-of-sample loss is computed as the sum over 10 different sets of randomly selected test data sets, here the improvements in the tails of the distributions are more marked than the in-sample performance.

\section{Discussion}
This paper proposes a novel simultaneous linear quantile regression model, named pyramid quantile regression (PQR), by using the quantile pyramids prior of \shortciteN{hjortw09} 
as a basis for building a flexible, nonparametric conditional density.

PQR avoids strong parametric assumptions about the conditional distributions, which adds great modelling flexibility and circumvents the need to make parametric assumptions about the distribution of the data. In addition, the model is parametrised in terms of the quantiles themselves, this is a natural way of modelling quantile regression and allows for easy interpretation and incorporation of prior information. For instance, one can centre the conditional quantile priors on chosen distributions based on prior knowledge. We considered centring it on the Normal distribution, and showed that this choice by default works well for a variety of cases, including mildly asymmetric densities. Additionally, PQR can be used for flexible extreme quantile modelling by centring the prior on an extreme distribution, as opposed to strictly requiring the data to follow the parametric assumption,  as is often the case in extreme value modelling. We illustrated this application in the modelling of extreme tropical cyclone winds in the US coast using pyramid prior centred on the Generalised Pareto Distribution (GPD). 
%PQR can also be extended to splines fitting, as illustrated here through the famous lidar dataset. 
The availability of an explicit expression for a likelihood affords easier extensions to more complex modelling. We have shown via simulation studies that PQR provides  robust estimates with small errors and great coverages properties.

We have demonstrated that the conditional quantiles implied by the linear regression model retains
posterior consistency. 
%Moreover, for fixed pyramid level $M$, quantile pyramids has posterior consistency to the least false parameters \textcolor{red}{(defined ?)}. 
%This suggests that inference may be based on fixed, finite $M$. In this context, simulation studies demonstrate that the proposed approach has small error and good coverage, overall outperforming its competitors, especially for extreme quantiles. 
%In addition, consistency to the truth is achieved by considering growing pyramid level $M$ with sample size,
Our experience with empirical studies also shows that $M$ does not need to be large to obtain reasonable results.
%, suggesting that the least false estimators are close to the truth.

%In summary, PQR is a novel technique for fitting single or multiple quantiles based on a flexible nonparametric model, providing
%\textcolor{red}{computational extensions for higher dimensions that allow for the full block updating can help mixing further.}

%#####

%In a similar way, the model we present here also interpolates the likelihood derived from quantiles. However, different interpolations and priors are considered, as well as very distinct modelling principles. While \shortciteN{fengch2015} aims to construct a good approximation to the likelihood and use it to estimate quantiles, our approach considers estimating quantiles from rough likelihoods. The model we propose emerges naturally under our prior specification, which in turn is based on the ingenious work of \shortciteN{hjortw09}. The framework allows us to fit a single quantile or as many specific levels of quantiles as we want.

\subsection*{Acknowledgements}%%
%%%%%%%%%%%%%%%%%%
%%%%%%%%%%%%%%%%%
TR is funded by CAPES Foundation via the Science Without Borders (BEX 0979/13-9). TR and YF are grateful to the Australian Research 
Council Centre of Excellence for Mathematical  and Statistical Frontiers for support.

\bibliographystyle{chicago}
\bibliography{pyramidQuantile.bbl}

\section*{Appendix}
For clarity we give the demonstrations for the case $P=1$, the generalization to $P>1$ with  ${\cal X}$ within the convex hull of the pyramid locations ${\bf x}^0,\dots,{\bf x}^P$ being straightforward. For $P=1$, without loss of generality, we suppose that  ${\bf x}^0=0$ and ${\bf x}^1=1$ so that, for  $0 < \tau <1$ and any $0\leq x \leq 1$,  
\begin{eqnarray*}
Q_{Y}(\tau|x) & = & (1-x) Q_{\tau}^0  + x Q_{\tau}^1 
%\label{linearity}
\end{eqnarray*}
where $Q_{\tau}^0$ and $Q_{\tau}^1$ are independent pyramid quantile processes. 
We have $Q_{\tau}^0=Q_{null}^0(Q^{0,unif}_{\tau})$ and $Q_{null}^1(Q^{0,unif}_{\tau})$ where $Q^{0,unif}_{\tau}$ and $Q^{1,unif}_{\tau}$ are independent pyramid quantile processes centered on the uniform distribution on $(0,1)$. 
We suppose that $Q_{\tau}^{0,unif}$ and $Q_{\tau}^{1,unif}$ are a.s. absolutely continuous and 
we suppose  that the two centring quantile functions $Q_{null}^{0}$ and $Q_{null}^{1}$ are also absolutely continuous. 
Thus $Q_{\tau}^0$ and $Q_{\tau}^1$ are   a.s. absolutely continuous and
we denote $q_0(\cdot)$ and $q_1(\cdot)$  the corresponding quantile density functions. 
Then, for any $0\leq x \leq 1$, the conditional quantile function $Q_{Y}(\tau|x)$ is also a.s. absolutely continuous with quantile density function $q_x(u)=(1-x)q_0(u)+xq_1(u)$.\\

\noindent {\bf Proof of Proposition \ref{KLsupport}}\\ 
%\textcolor{red}{do we need to define $q_0$ and $q_1$?}\\
We first show that conditions  similar to (B) and (C) are also true at any $x \in (0,1)$ : 
\begin{description}
\item[$({B}_x)$]{\it  for all $\delta>0$ there exists an $\epsilon>0$ such that, $\forall \, x  \in (0,1)$,
$$\int \ln\frac{ q_x^*(\tau_\epsilon(u))}{q_x^*(u)} du < \delta$$ 
for any function $\tau_{\epsilon}(u)$ from $[0,1]$ to $[0,1]$  for which $\max_u |\tau_{\epsilon}(u)-u|<\epsilon$.   }\\
We use the  log sum inequality and see that, $\forall \, x \in (0,1)$,
\begin{eqnarray*}
\int \ln\frac {q_x^*(\tau_\epsilon(u))}{q_x^*(u)} du & = & \int \ln\frac {(1-x)q_0^*(\tau_\epsilon(u))+xq_1^*(\tau_\epsilon(u))}{(1-x)q_0^*(u)+xq_1^*(u)} du\\
 & \leq & \int \frac{1}{(1-x)q_0^*(\tau_\epsilon(u))+xq_1^*(\tau_\epsilon(u))}\left\{ (1-x)q_0^*(\tau_\epsilon(u)) \ln \frac{q_0^*(\tau_\epsilon(u))}{q_0^*(u)}  \right.\\
  & & \qquad \qquad\qquad  \left.  +xq_1^*(\tau_\epsilon(u)) \ln \frac{q_1^*(\tau_\epsilon(u))}{q_1^*(u)}   \right\} du \\
  & \leq &\int  \ln \frac{q_0^*(\tau_\epsilon(u))}{q_0^*(u)}du + \int \ln \frac{q_1^*(\tau_\epsilon(u))}{q_1^*(u)}du
\end{eqnarray*}
and by using condition (B) we get the result.
\item[$({C}_x)$]{\it  $\forall \, x \in (0,1)$ the density $f_x$ is bounded by some $K<\infty$.}\\
Under the condition (C) $f_0$ and $f_1$ are bounded by some finite $K_0$ and   $K_1$. Since $f_x(\cdot)=1/q_x(F_x(\cdot))$   we have, $\forall \, x \in (0,1)$, 
%by using $f_x(\cdot)=1/q_x(F_x(\cdot))$,
\begin{eqnarray*}
q_x(\cdot)=(1-x)q_0(\cdot)+xq_1(\cdot) & > & (1-x)\frac{1}{K_0}+x \frac{1}{K_1}
\end{eqnarray*}
thus, $\forall \, x \, \in (0,1)$, 
\begin{eqnarray*}
f_x (\cdot) & < & \left\{  (1-x)\frac{1}{K_0}+x \frac{1}{K_1} \right\}^{-1}<\infty.
\end{eqnarray*}
%and this is verified.
\end{description}
Once these properties are stated we can follow step by step the lines of  the proof of Proposition 3.1 in \shortciteN{hjortw09}. For any $x$ in $(0,1)$, by using the change of variable $u=F_x^*(y)$, the Kullback-Leibler divergence between $f_x^*$ and $f_x$ can be decomposed as %the sum of two terms
\begin{eqnarray*}
\int f_x^*(y) \ln\frac{f_x^*(y)}{f_x(y)} dy & = & \int \ln \frac{q_x (\tau_x(u))}{q_x^*(u)}du\\
& = &  \int \ln \frac{q_x (\tau_x(u))}{q_x ^*(\tau_x(u))}du + \int \ln \frac{q_x ^*(\tau_x(u))}{q_x^*(u)}du
\end{eqnarray*}
where $\tau_x(u)=F_x(Q_x^*(u))$. 
Proceeding as in \shortciteN{hjortw09}, and using conditions $({B}_x)$ and $({C}_x)$, the first term in this sum is smaller than any arbitrary  positive value with positive prior probability mass  if, for any $\epsilon>0$, the prior puts positive probability mass on $\{Q_x:\max_u|\lambda_x(u)-u|<\epsilon\}$ where $\lambda_x(u)=F_x^*(Q_x(u))$. 
%Using the property $({C}_x)$ the first term in this sum is smaller than any arbitrary  positive value with positive prior probability mass provided that, for any $\epsilon>0$, the prior puts positive probability mass on $\{Q_x:\max_u|\lambda_x(u)-u|<\epsilon\}$ where $\lambda_x(u)=F_x^*(Q_x(u))$. 
To prove that this sufficient condition is true note that we have, $\forall u \in (0,1)$,
\begin{eqnarray*}
|Q_x(u)-Q_x^*(u)| & = &  |(1-x)(Q^0_{\tau}(u)-Q^{*0}_{\tau}(u))+x (Q^1_{\tau}(u)-Q^{*1}_{\tau}(u)) |\\
 & \leq & |Q^0_{\tau}(u)-Q^{*0}_{\tau}(u)|+|Q^1_{\tau}(u)-Q^{*1}_{\tau}(u) |.
\end{eqnarray*}
Now, from condition (A), the prior puts positive probability mass on $\{Q^{0,unif}_{\tau}:\max_u|Q^{0,unif}_{\tau}(u)-Q^{*0,unif}_{\tau}(u)|<\theta^0\}$ and $\{Q^{1,unif}_{\tau}:\max_u|Q^{1,unif}_{\tau}(u)-Q^{*1,unif}_{\tau}(u)|<\theta^1\}$ for any positive $\theta^0$ and $\theta^1$. 
Thus, from the absolute continuity of $Q_{null}^0$ and $Q_{null}^1$, the prior puts positive probability mass on $\{Q^{0}_{\tau}:\max_u|Q^{0}_{\tau}(u)-Q^{*0}_{\tau}(u)|<\delta^0\}$ and $\{Q^{1}_{\tau}:\max_u|Q^{1}_{\tau}(u)-Q^{*1}_{\tau}(u)|<\delta^1\}$ for any positive $\delta^0$ and $\delta^1$ and so, using the preceding inequality, puts positive   probability mass on $\{Q_x:\max_u|Q_x(u)-Q_x^*(u)|<\delta\}$ for any positive $\delta$. By using the absolute continuity of $F_x^*$ we finally get that, for any positive $\epsilon$,  the prior puts positive probability mass on $\{Q_x:\max_u|\lambda_x(u)-u|<\epsilon\}$.

For the second term in the sum we use again the consequence of condition (A): the prior puts positive probability mass on $\{Q_x:\max_u|Q_x(u)-Q_x^*(u)|<\delta\}$ for any positive $\delta$ then, using the absolute continuity of $F_x$, puts positive probability mass on $\{F_x:\max_u|\tau_x(u)-u|<\epsilon\}$ for any $\epsilon>0$. Hence, using the property $({B}_x)$, this term is also bounded by any positive real with positive probability and finally we know that the prior put positive probability mass on $\{f_x: d_{KL}(f_x^*,f_x) < \epsilon\}$.

To complete the proof note that this result is true for any $x\in(0,1)$ and we have, for any $\epsilon>0$, 
$$
\Pi \left( \{f:   \forall \, 0\leq x \leq 1 \, d_{KL}(f_x^*,f_x) < \epsilon\}   \right)>0.
$$
Since 
\begin{eqnarray*}
d_{KL}(f^*,f) & = &  \int   d_{KL}(f_x^*,f_x)  {f_X}(x) dx
\end{eqnarray*}
%\textcolor{red}{is it better to remove $(y)$ in the above?}\\
we get the desired result. $\square$\\ 

\noindent {\bf Proof of Proposition \ref{HLconsistency}}\\ 
We have just to follow the steps of the proof of proposition 7.1 in \shortciteN{hjortw09} and to note that the Hellinger distance is given by 
\begin{eqnarray*}
d_h^2(f^*,f) & = &  \int d_H(f_x^*,f_x) {f_X}(x) dx
\end{eqnarray*} 
and that, if  $q_{kj}$, $j=1,..., 2^{M_n}-1$, $k=0,1$ are the quantile sampled by $\Pi_{M_n}$, if for $j=1,..., 2^{M_n}-1$, if we have both $|q_{0j}-q_{0j}^*|<\epsilon$ and $|q_{1j}-q_{1j}^*|<\epsilon$ then, $\forall \, x \in (0,1)$,
$$
|q_{xj}-q_{xj}^*|  \leq  (1-x)|q_{0j}-q_{0j}^*|+x|q_{1j}-q_{1j}^*| \leq  \epsilon.
$$
It turns out that, for a given $\delta>0$, $\forall \, x \in (0,1)$, there exits $\epsilon>0$ such that if $|q_{0j}-q_{0j}^*|<\epsilon$ and $|q_{1j}-q_{1j}^*|<\epsilon$ then $d_H^2(f^*,f) <\delta$. Once this is stated the rest of the proof of \shortciteN{hjortw09} applies. $\square$\\

\end{document}